\documentclass[a4paper,11pt,numreference,epsfig,cite]{article}

\usepackage[total={5in, 9.5in}]{geometry}
\usepackage[latin1]{inputenc}
\usepackage{textcomp}
\usepackage{graphicx}
\usepackage{xcolor}
\usepackage{amsmath}

\newcommand{\stt}{\small\tt}

\newcommand{\Abstract}[1]{\begin{center}{\stt ABSTRACT}\end{center}{#1}
\vskip 0.4in}

\title{${\rm \overline{B}}_{d,s}\to \rm K^{*0}\overline{K}^{*0}$ decays, a serious problem for the
Standard Model
 \protect\\}

\author{R. Aleksan$^1$, L. Oliver$^2$ \\
\footnotesize $^1$IRFU, CEA, Universit\'e Paris-Saclay, 91191 Gif-sur-Yvette cedex, France \\
\footnotesize $^2$IJCLab, P\^ole Th\'eorie, CNRS/IN2P3 et Universit\'e Paris-Saclay, \\
\footnotesize B\^at. 210, 91405 0rsay, France }

\def\BdKK{{\rm \overline{B}}_d \to {\rm K^{*0} \overline{K}^{*0}} }
\def\BsKK{{\rm \overline{B}}_s \to {\rm K^{*0} \overline{K}^{*0}} }
\def\BdsKK{{\rm \overline{B}}_{d,s} \to {\rm K^{*0} \overline{K}^{*0}} }
\def\Bsphiphi{{\rm \overline{B}}_s \to\phi\phi }
\begin{document}

\maketitle

\Abstract{\normalsize \baselineskip 22pt
We underline the theoretical and experimental interest of the vector-vector penguin decays $\BdsKK$ for which the data show a strong U-spin violation. Indeed, with the latest LHCb data one has for these two modes very different longitudinal polarization fractions, $f_L^{exp}(\BsKK ) = 0.24 \pm 0.04$ and $f_L^{exp}(\BdKK ) = 0.74 \pm 0.05$. This feature is very striking because both modes are related by the exchange $s \leftrightarrow d$, i.e. they are related by U-spin symmetry as pointed out in earlier work by other authors. We illustrate this phenomenon by computing different observables for these modes within the QCD Factorization scheme, and we find, as expected, rather close values for the longitudinal fractions, with central values $f_L^{th}(\BsKK ) \simeq 0.42$ and $f_L^{th}(\BdKK ) \simeq 0.49$ and rather small errors. Furthermore, due to the $V-A$ nature of the weak currents and the heavy quark limit, one expects $(f_\parallel/f_\perp)_{th}\simeq 1$, which in the $\rm B_s$ case is in contradiction with the data $(f_\parallel/f_\perp)_{\rm exp}\simeq 0.44\pm 0.06$.}

% ================================ new section ===============================

\section{Introduction}

In the absence of clear indications of physics beyond the Standard Model (SM), one may study in detail the rare decays of particles to search for deviations from the SM predictions. In this regard, B-meson decays to two vector particles are a very rich source of observables, which can exibit signs of New Physics (NP). It has been shown that the decay $\Bsphiphi$ is a good candidate in order to search for evidence of NP~\cite{AO:1}, in particular at FCC-ee~\cite{fccee:1,fccee:2,fccee:3} where high sensitivities can be obtained. This can be achieved by searching for CP violating effects in time-dependent studies while almost no effect is expected in the Standard Model. Similar studies can be carried out with the decays  $\BsKK$, with which sensitivities to NP can be improved further. In the present paper, we study this latter decay together with the decay $\BdKK$, which is related to the former through U-spin symetry, and we show that it behaves in a way which is theoretically unexpected using QCD Factorization, thus requiring particular attention. 

% ================================ new section ===============================

\section{${\rm B}\to V_1V_2$ experimental data}

\noindent We summarize in Table~\ref{tab:B_decays} the available data on $\rm B$ decays to 2 light vector particles as given by the Particle Data Group (PDG)~\cite{pdg:1}.

%\newgeometry{hmargin={1cm}}

\begin{table}[htb]
\centering
\small

$$ \hspace{-2.5cm}
\begin{tabular}{lcccccc}

$\displaystyle {\mathrm {{B}\ decay}} $ &
$\displaystyle {\mathrm {Br} (\times10^{-6})} $ & 
$\displaystyle {f_L} $ &
$\displaystyle {f_\parallel} $  &
$\displaystyle {f_\perp} $ &
$\displaystyle {f_\parallel - f_\perp} $ &
$\displaystyle A_{\rm CP} $ \\
\hline \hline

\textcolor{red} {\underline{ {\boldmath $\overline{B}_s $} modes}} \\ 

$\displaystyle \overline{B}_s\to \phi\rho^0$ &
$\displaystyle {\mathrm {0.27\pm0.08}}$ &
$\displaystyle {\mathrm {-}} $ &
$\displaystyle {\mathrm {-}} $ &
$\displaystyle {\mathrm {-}} $ &
$\displaystyle {\mathrm {-}} $ &
$\displaystyle {\mathrm {-}} $ \\

$\displaystyle \overline{B}_s\to \phi\phi$ &
$\displaystyle {\mathrm {18.5\pm1.4}}$ &
$\displaystyle {\mathrm {0.378 \pm 0.013}} $ &
$\displaystyle {\mathrm {{\it 0.330\pm 0.016}}} $ &
$\displaystyle {\mathrm {0.292 \pm 0.009}} $ &
$\displaystyle {\mathrm {0.038 \pm {0.022}}} $ &
$\displaystyle {\mathrm {-}} $ 
\\

$\displaystyle \overline{B}_s\to \overline{K}^{*0}{K}^{*0}$ &
$\displaystyle {\mathrm {11.1\pm2.7}}$ &
$\displaystyle {\mathrm {0.24 \pm 0.04}} $ &
$\displaystyle {\mathrm {0.297\pm 0.049}}^\ast $ &
$\displaystyle {\mathrm {0.38 \pm 0.12}}^\ast $ & 
$\displaystyle {\mathrm {-0.08 \pm {0.13}}}^\ast $ &
$\displaystyle {\mathrm {-}} $ \\

$\displaystyle \overline{B}_s\to \phi{K}^{*0}$ &
$\displaystyle {\mathrm {1.14\pm0.30}}$ &
$\displaystyle {\mathrm {0.51 \pm 0.17}} $ &
$\displaystyle {\mathrm {0.21\pm 0.11}} $ &
$\displaystyle {\mathrm {{\it 0.28\pm 0.20 }}} $ &
$\displaystyle {\mathrm {-0.07 \pm {0.28}}} $ &
$\displaystyle {\mathrm {-}} $ \\

\textcolor{red} {\underline{ {\boldmath $\overline{B}_d $} modes}}\\ 

$\displaystyle \overline{B}^0\to \omega \omega$ &
$\displaystyle {\mathrm {1.2\pm0.4}}$ &
$\displaystyle {\mathrm {-}} $ &
$\displaystyle {\mathrm {-}} $ &
$\displaystyle {\mathrm {-}} $ &
$\displaystyle {\mathrm {-}} $ &
$\displaystyle {\mathrm {-}} $ \\

$\displaystyle \overline{B}^0\to \rho^{+}\rho^{-}$ &
$\displaystyle {\mathrm {27.7\pm1.9}}$ &
$\displaystyle {\mathrm {0.990 \pm ^{0.021}_{0.019}}} $ &
$\displaystyle {\mathrm {-}} $ &
$\displaystyle {\mathrm {-}} $ &
$\displaystyle {\mathrm {-}} $ &
$\displaystyle {\mathrm {-}} $ \\

$\displaystyle \overline{B}^0\to \rho^{0}\rho^{0}$ &
$\displaystyle {\mathrm {0.96\pm0.15}}$ &
$\displaystyle {\mathrm {0.71 \pm ^{0.08}_{0.09}}} $ &
$\displaystyle {\mathrm {-}} $ &
$\displaystyle {\mathrm {-}} $ &
$\displaystyle {\mathrm {-}} $ &
$\displaystyle {\mathrm {-}} $ \\

$\displaystyle \overline{B}^0\to \rho^+ K^{*-}$ &
$\displaystyle {\mathrm {10.3\pm2.6}}$ &
$\displaystyle {\mathrm {0.38 \pm 0.13}} $ &
$\displaystyle {\mathrm {-}} $ &
$\displaystyle {\mathrm {-}} $ &
$\displaystyle {\mathrm {-}} $ &
$\displaystyle {\mathrm {0.21\pm  0.15}} $ \\

$\displaystyle \overline{B}^0\to \rho^0 \overline{K}^{*0}$ &
$\displaystyle {\mathrm {3.9\pm1.3}}$ &
$\displaystyle {\mathrm {0.173 \pm 0.026}} $ &
$\displaystyle {\mathrm {{\it 0.426\pm 0.048}}} $ &
$\displaystyle {\mathrm {0.401\pm 0.040}} $ &
$\displaystyle {\mathrm {0.025 \pm {0.084}}} $ &
$\displaystyle {\mathrm {-0.16 \pm {0.06}}} $\\

$\displaystyle \overline{B}^0\to \omega \overline{K}^{*0}$ &
$\displaystyle {\mathrm {2.0\pm0.5}}$ &
$\displaystyle {\mathrm {0.69 \pm 0.11}} $ &
$\displaystyle {\mathrm {{\it 0.21\pm 0.17}}} $ &
$\displaystyle {\mathrm {0.10\pm 0.13}} $ &
$\displaystyle {\mathrm {0.11 \pm {0.28}}} $ &
$\displaystyle {\mathrm {0.45\pm 0.25}} $ 
\\

$\displaystyle \overline{B}^0\to \overline{K}^{*0}\phi$ &
$\displaystyle {\mathrm {10.0\pm0.5}}$ &
$\displaystyle {\mathrm {0.497 \pm 0.017}} $ &
$\displaystyle {\mathrm {{\it 0.279\pm 0.023}}} $ &
$\displaystyle {\mathrm {0.224\pm 0.015}} $ &
$\displaystyle {\mathrm {0.055 \pm {0.034}}} $ &
$\displaystyle {\mathrm {0.00\pm 0.04}} $ \\

$\displaystyle \overline{B}^0\to K^{*0}\overline{K}^{*0}$ &
$\displaystyle {\mathrm {0.83\pm0.24}}$ &
$\displaystyle {\mathrm {0.74 \pm 0.05}} $ &
$\displaystyle {\mathrm {-}}^\ast $ &
$\displaystyle {\mathrm {-}} ^\ast$ &
$\displaystyle {\mathrm {-}}^\ast $ & 
$\displaystyle {\mathrm {-}} $ \\ 
\hline \hline

\textcolor{red} {\underline{ {\boldmath $\overline{B}_u $} modes}}\\ 

$\displaystyle B^-\to \omega {K}^{*-}$ &
$\displaystyle {\mathrm {<7.4 }}$ &
$\displaystyle {\mathrm {0.41 \pm 0.18}} $ &
$\displaystyle {\mathrm {-}} $ &
$\displaystyle {\mathrm {-}} $ &
$\displaystyle {\mathrm {-}} $ &
$\displaystyle {\mathrm {0.29\pm 0.35}} $ \\

$\displaystyle B^-\to \omega{\rho}^{-}$ &
$\displaystyle {\mathrm {15.9\pm2.1}}$ &
$\displaystyle {\mathrm {0.90 \pm 0.06}} $ &
$\displaystyle {\mathrm {-}} $ &
$\displaystyle {\mathrm {-}} $ &
$\displaystyle {\mathrm {-}} $ &
$\displaystyle {\mathrm {-0.20\pm 0.09}} $ \\

$\displaystyle B^-\to \rho^0{\rho}^{-}$ &
$\displaystyle {\mathrm {24.0\pm1.9}}$ &
$\displaystyle {\mathrm {0.950 \pm 0.016}} $ &
$\displaystyle {\mathrm {-}} $ &
$\displaystyle {\mathrm {-}} $ &
$\displaystyle {\mathrm {-}} $ &
$\displaystyle {\mathrm {-0.05\pm 0.05}} $ \\

$\displaystyle B^-\to \rho^0{K}^{*-}$ &
$\displaystyle {\mathrm {4.6\pm1.1}}$ &
$\displaystyle {\mathrm {0.78 \pm 0.12}} $ &
$\displaystyle {\mathrm {-}} $ &
$\displaystyle {\mathrm {-}} $ &
$\displaystyle {\mathrm {-}} $ &
$\displaystyle {\mathrm {0.31\pm 0.13}} $ \\

$\displaystyle B^-\to \rho^-\overline{K}^{*0}$ &
$\displaystyle {\mathrm {9.2\pm1.5}}$ &
$\displaystyle {\mathrm {0.48 \pm 0.08}} $ &
$\displaystyle {\mathrm {-}} $ &
$\displaystyle {\mathrm {-}} $ &
$\displaystyle {\mathrm {-}} $ &
$\displaystyle {\mathrm {-0.01\pm 0.16}} $ \\

$\displaystyle B^-\to K^{*-}\phi $ &
$\displaystyle {\mathrm {10.0\pm2}}$ &
$\displaystyle {\mathrm {0.50 \pm 0.05}} $ &
$\displaystyle {\mathrm {{\it 0.30\pm 0.07}}} $ &
$\displaystyle {\mathrm {0.20\pm 0.05}} $ &
$\displaystyle {\mathrm {0.10 \pm {0.12}}} $ &
$\displaystyle {\mathrm {-0.01\pm 0.08}} $ \\

$\displaystyle B^-\to K^{*-}{K}^{*0}$ &
$\displaystyle {\mathrm {0.91\pm0.29}}$ &
$\displaystyle {\mathrm {0.82 \pm ^{0.15}_{0.21}}} $ &
$\displaystyle {\mathrm {-}} $ &
$\displaystyle {\mathrm {-}} $ &
$\displaystyle {\mathrm {-}} $ &
$\displaystyle {\mathrm {-}} $ \\
	
\hline\hline

\hline

\end{tabular}    $$

\caption{\small \label{tab:B_decays}  B-meson branching fractions, $f_L$, $f_\parallel$, $f_\perp$ and $A_{CP}$  for some selected $V_1V_2$ modes from the PDG~\cite{pdg:1}. Statistical and systematic errors have been added in quadrature. The values in italic are not measured directly but are deduced from $f_L+ f_\parallel +f_\perp = 1$. $^\ast$ For these modes, LHCb has made more precise measurements not included by PDG and quoted below in the text.}
\end{table}

%\restoregeometry

\noindent In this Table~\ref{tab:B_decays}, the Branching fractions and polarizations fractions are CP averages, i.e.
\begin{equation}
\small
\begin{array}{lclcccc} 
\mathrm{Br}  & =  & 	\frac{1}{2}[\mathrm{Br}(B_{d,s} \to f) + \mathrm{Br}(\overline{B}_{d,s} \to \overline{f})]\\
\\
f_{L,\parallel,\perp}  & =  & 	\frac{1}{2}[f_{L,\parallel,\perp}(B_{d,s} \to f) + f_{L,\parallel,\perp}(\overline{B}_{d,s} \to \overline{f})]\\
\\
A_{CP} & = & \frac{ \mathrm{Br}(\overline{B}_{d,s} \to \overline{f})) - \mathrm{Br}({B}_{d,s} \to {f})}{{ \mathrm{Br}(\overline{B}_{d,s} \to \overline{f})) + \mathrm{Br}({B}_{d,s} \to {f})}} \\
\\
A_{CP}^h & = & \frac{ f_h(\overline{B}_{d,s} \to \overline{f}) - f_h({B}_{d,s} \to {f})}{ f_h(\overline{B}_{d,s} \to \overline{f}) + f_h({B}_{d,s} \to {f})} \ \ \mathrm{with} \ h = L,\parallel ,\perp
\end{array}
\label{eq:CPaverage}
\end{equation}

\vskip 10pt
\noindent It is important to note that LHCb has measured $f_L,\ f_\parallel$ and $ f_\perp$ for $\BsKK$~\cite{lhcb:1} more precisely and finds
\begin{equation}
\begin{array}{lclcccc} 
f_L =  0.240\pm 0.040 & ,  & f_\parallel = 0.234\pm 0.027& , & f_\perp = 0.526\pm 0.037
\end{array}
\label{eq:LHCb_Bs}
\end{equation}

\noindent leading to  $\Delta_{f_{\parallel,\perp}}=f_\parallel -f_\perp = -0.292 \pm 0.046$, i.e $\neq 0$ at 6.4 $\sigma$. LHCb has also measured $f_L$, $f_\parallel$ and $f_\perp$ for $\BdKK$~\cite{lhcb:1} and finds 

\begin{equation}
\begin{array}{lclcccc} 
f_L = 0.724\pm 0.053 & , & f_\parallel = 0.116\pm 0.035 & , & f_\perp = 0.160\pm 0.046
\end{array}
\label{eq:LHCb_Bd}
\end{equation}
\noindent We note that, for this mode, $f_\parallel -f_\perp = -0.044 \pm 0.058$, i.e. compatible with  $\Delta_{f_{\parallel,\perp}}=0$. In summary, using the LHCb data, one has two striking features, which will be discussed further :

\begin{equation}
\left(\frac{f_{\parallel}^{\BsKK}} {f_\perp ^{\BsKK}}\right)_{\mathrm{LHCb}}  =  0.44  \pm  0.06 \\
\label{eq:f_Tratio}
\end{equation}
\begin{equation}
\left(\frac{f_{L}^{\BdKK}} {f_L ^{\BsKK}}\right)_{\mathrm{LHCb}}  = 3.02  \pm  0.55 
\label{eq:f_Lratio}
\end{equation}
\section{Expectations using Naive Factorization}
\vskip 10pt
\noindent A detailed theoretical study of $\overline{B}\to V_1V_2$ has been carried out~\cite{AO:2} using QCD Factorization. In the following, we summarize the theoretical expectations using the values of the parameters shown in the Tables~\ref{tab:paramCKM}, \ref{tab:paramB} and \ref{tab:param}.
\vskip 10pt
\noindent In $\rm \overline{B}\to V_1V_2$ decays, one is dealing with 3 helicity amplitudes.
\begin{equation}
\begin{array}{lclcccc} 
{\cal \overline{A}}_0 & = & A[\overline{B}\to V_1(0)V_2(0)] & , & {\cal \overline{A}}_\pm & = &   A[\overline{B}\to V_1(\pm)V_2(\pm)]    \\
\end{array}
\label{eq:A}
\end{equation}

\vskip 10pt
\noindent Moving from the helicity representation to the transversity one, one gets :
\begin{equation}
\begin{array}{lclcccc} 
{\cal  \overline{A}}_L& = & {\cal \overline{A}}_0 \\
\\
{\cal \overline{A}}_\parallel& = & \frac{{\cal \overline{A}}_+ +  {\cal \overline{A}}_-}{\sqrt{2}} & , & {\cal \overline{A}}_\perp& = & \frac{{\cal \overline{A}}_+ -  {\cal \overline{A}}_-}{\sqrt{2}}  \\
\end{array}
\label{eq:A_trans}
\end{equation}

\noindent with the corresponding transversity rate fractions $f_L$, $f_\parallel$ and $f_\perp$ satisfying
\begin{equation}
\begin{array}{lclcccc} 
f_L+f_\parallel +f_\perp = 1  \\
\end{array}
\label{eq:f_eq}
\end{equation}

\vskip 10pt
\noindent From Beneke et al. ~\cite{BRY:1}, one has :
\begin{equation}
\begin{array}{cccc} 
{\cal \overline{A}}_0^{V_1V_2} & =  & i\frac{G_F}{\sqrt{2}} m_B^2f_{V_2} A_0^{B\to V_1}(0)\\
\\
{\cal \overline{A}}_\pm^{V_1V_2} & =  & i\frac{G_F}{\sqrt{2}} m_Bm_{V_2}f_{V_2} F_\pm^{B\to V_1}(0)\\
\end{array}
\label{eq:Amplitude}
\end{equation}
\vskip 10pt
\noindent with
\begin{equation}
\begin{array}{cccc} 
F_\pm^{B\to V_1}(q^2) & \equiv  & \left(1+\frac{m_1}{{m_B}}\right) A_1^{B\to V_1}(q^2)\mp \left(1-\frac{m_1}{{m_B}}\right) V^{B\to V_1}(q^2)\\
\end{array}
\label{eq:FF}
\end{equation}

\noindent Since the final quark is dominantly left-handed because of the $V-A$ structure of the Standard Model, heavy quark symmetry implies the hierarchy
\begin{equation}
\begin{array}{cccc} 
{\cal \overline{A}}_L : {\cal \overline{A}}_- : {\cal \overline{A}}_+ & = & 1 : \frac{\Lambda_{QCD}}{m_b} : \left( \frac{\Lambda_{QCD}}{m_b} \right)^2\\
\end{array}
\label{eq:hierarchy}
\end{equation}

\vskip 10pt
\noindent The transverse amplitude ${\cal \overline{A}_-}$ is suppressed by factor $m_{V_2}/m_b$ relative to ${\cal \overline{A}}_L$, and the axial and vector contributions to ${\cal \overline{A}}_+$ cancel out in the heavy quark limit and large recoil energy for the light mesons. Indeed, in this latter limit, the $A_1$ and $V$ form factors are related and one has~\cite{BF:1}
\begin{equation}
\begin{array}{cccc} 
\frac{m_B}{m_B+m_V}V(q^2) & =  & \frac{m_B+m_V}{2E}A_1(q^2)\\
\\
\mathrm{with} \ E & = & (m_B^2+m_V^2 -q^2)/2m_B
\end{array}
\label{eq:FFactors}
\end{equation}
\vskip 10pt
\noindent At $q^2=0$ , one gets $2Em_B=m_B^2+m_V^2$ and therefore in the limit $m_V << m_B$,
\begin{equation}
\begin{array}{cccccc} 
\frac{V(0)}{A_1(0)} & \simeq  & 1\\
\end{array}
\label{eq:FFratio}
\end{equation}
\vskip 10pt
\noindent giving in the heavy quark limit and large recoil energy for the light meson,
\vskip 10pt 
\begin{equation}
\begin{array}{cccccc} 
{\cal \overline{A}}_+\simeq 0
\end{array}
\label{eq:A=0}
\end{equation}
\noindent and one then has 
\begin{equation}
\begin{array}{cccc} 
{\cal \overline{A}}_\parallel \simeq   -{\cal \overline{A}}_\perp& \simeq & \frac{{\cal \overline{A}}_-}{\sqrt{2}}  \\
\\
|{\cal \overline{A}}_\parallel|^2 + |{\cal \overline{A}}_\perp|^2& \simeq & |{\cal \overline{A}}_-|^2 \equiv |{\cal \overline{A}}_T|^2\\
\end{array}
\label{eq:F}
\end{equation}
\vskip 10pt
\noindent With the hierachy above and the limit ${\cal \overline{A}}_+ \simeq 0$, one gets
\begin{equation}
\begin{array}{cccc} 
f_\parallel & \simeq&  f_\perp
\end{array}
\label{eq:F=F}
\end{equation}
\vskip 10pt
\noindent Accordingly, we are left with only one transverse form factor ${F_T^{B_{d,s}\to x}(0)} \equiv {F_-^{B_{d,s}\to x}(0)} $.
As can be seen in Table~\ref{tab:B_decays}, the experiment is in agreement with equations~(\ref{eq:F}) within errors, with the noticeable exception of $\BsKK$ if one uses the LHCb measurements of formula~(\ref{eq:LHCb_Bs}). This latter mode is discussed further in this note.

\vskip 10pt
\noindent In most Penguin dominated decays such as $\Bsphiphi$, as we have exposed in detail in \cite{AO:1}, although the equality $f_\parallel \simeq f_\perp$ is approximately satisfied by the data, the longitudinal  fraction $f_L$ is experimentally much smaller, of the order of $f_{\parallel,\perp}$, a feature that is qualitatively described within the QCD Factorization (QCDF) scheme \cite{BBNS-01, BBNS-99, BBNS-00}, in papers devoted to the $B$ decays to two vector mesons $\rm \overline{B}_{d,s} \to V_1 V_2$  \cite{BRY:1, KAGAN, CHENG-YANG, BUCHALLA}.

% ================================ new section ===============================

\section{$\BdsKK$ decays using QCD Factorization}

We will now focus on the decays $\BdsKK$. The important point that we want to underline is that these two decays are related by $U$-spin symmetry $d \leftrightarrow s$, as shown in all generality in Fig. 1 and already pointed out in e.g.~\cite{lhcb:1,CPS:1,DMV:1,DMV:2,Alguero,Biswas}. Therefore, the observables $f_{L,\parallel,\perp}$ in both decays should be equal in the limit of exact $SU(3)$ symmetry.\par
%%%%%%%%%%%%%%%%%%%% Figure Feynman diagram Bs to DsK  %%%%%%%%%%%%%%%%%%%
\begin{figure}[hbt]
\vfill
\begin{center}

%\vskip 5cm
\includegraphics[width=0.38\textwidth]{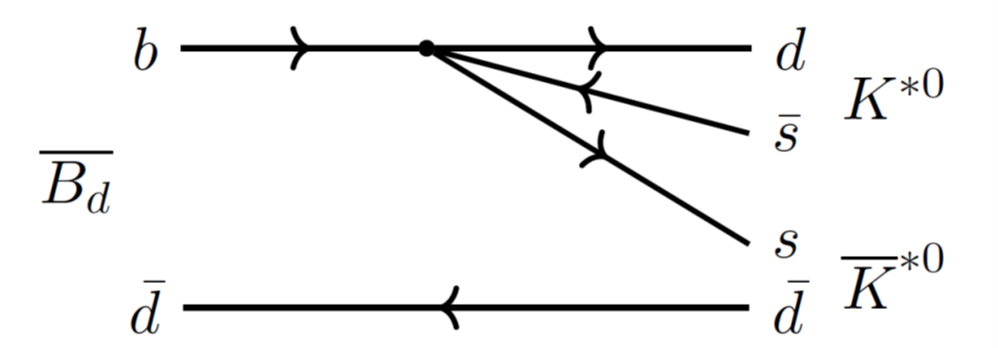}
\includegraphics[width=0.58\textwidth]{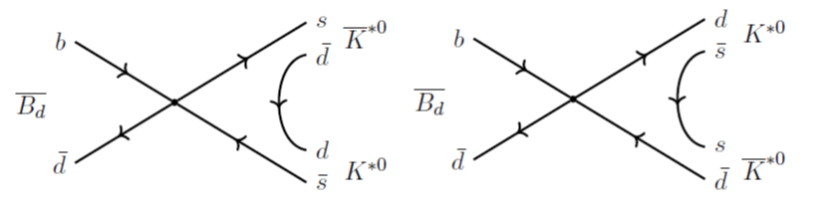}
\includegraphics[width=0.38\textwidth]{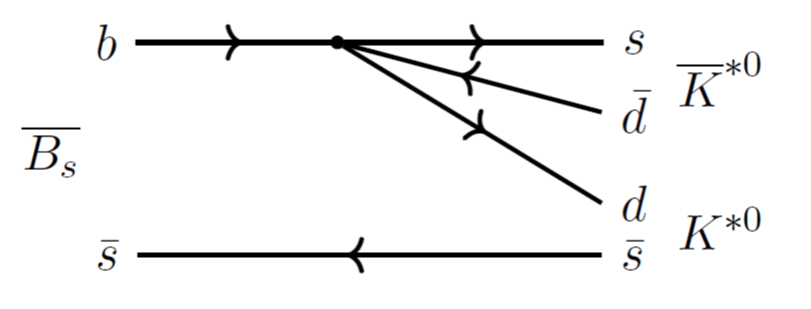}
\includegraphics[width=0.58\textwidth]{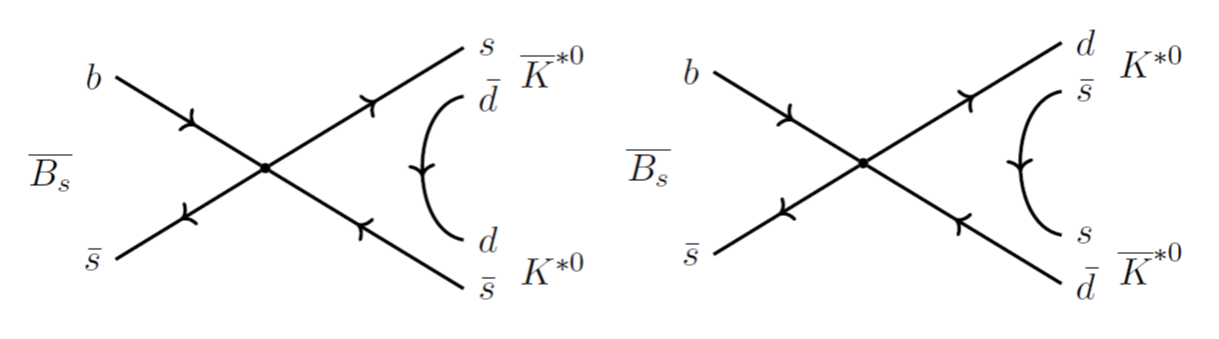}
\caption{\label{fig:U-spin}  $U$-spin symmetry relates the modes $\BsKK$ and $\BdKK$.}
\end{center}
\vfill

\end{figure}
%%%%%%%%%%%%%%%%%%%%%%%%% End Figure %%%%%%%%%%%%%%%%%%%%%%%%%
\vskip 10pt
\noindent However, the experimental results of LHCb \cite{lhcb:1} noted in~(\ref{eq:LHCb_Bs}) and~(\ref{eq:LHCb_Bd}) strongly contradicts this expectation.

\vskip 10pt
\noindent In the limit of Naive Factorization, the amplitudes for the modes $\BsKK$ and $\BdKK$ are respectively governed by the products of CKM factors
\begin{equation}
\label{1e} 
\begin{array}{cccccc} 
A(\BsKK) \sim \lambda'_t = V_{tb}V^*_{ts} = - \lambda'_u - \lambda'_c \\
A(\BdKK) \sim \lambda_t = V_{tb}V^*_{td} =  - \lambda_u - \lambda_c
\end{array} 
\end{equation}

\noindent and the expected ratio of rates is of the order of magnitude
\begin{equation}
\label{2e} 
{{\rm Br}(\BdKK) \over {\rm Br}(\BsKK)} \simeq 0.046
\end{equation}

\noindent i.e. consistent within errors with the branching ratios of Table 1,
\begin{equation}
\label{3e} 
\left[{{\rm Br}(\BdKK) \over {\rm Br}(\BsKK)}\right]_{\rm exp} = 0.075\pm 0.028
\end{equation}

\noindent We conclude that, for the time being, it does not seem to be a contradiction between theory and experiment for the rates.\par

\vskip 10pt
\noindent Of course, the hadronic parameters like form factors and decay constants are not the same for both modes. This feature will introduce some $U$-spin violation. To have a reasonable estimation of this $U$-spin violation, we will now compute the observables for both decays within QCD Factorization.\par

\vskip 10pt
\noindent Following the notation of \cite{AO:1}, the amplitudes of both modes write
$$A(\BsKK,h) = \sum_{p=u,c} \lambda '_p S^{p,h} A^h(\BsKK)$$
\begin{equation}
\label{4e} 
+\ (\lambda '_u + \lambda '_c) T^h B^h(\BsKK)
\end{equation}

$$A(\BdKK,h) = \sum_{p=u,c} \lambda_p S^{p,h} A^h(\BdKK)$$
\begin{equation}
\label{5e}  +\ (\lambda _u + \lambda _c) T^h B^h(\BdKK)
\end{equation}

\vskip 10pt

\noindent where the first terms describe the production-emission decays and the second ones correspond to the annihilation, as described in Fig. 1 and the CKM factors are given in Table~\ref{tab:paramCKM},

\begin{table}[htb]
\small
\centering
$$ \begin{tabular}{lccccccc}
\hline

$\displaystyle \lambda_u $  &
$\displaystyle V_{ub}V_{ud}^*$ &
$\displaystyle (0.001286\pm 0.000152) - (0.003268 \pm 0.000115)i$ &

\\ 
$\displaystyle \lambda_c $  &
$\displaystyle V_{cb}V_{cd}^*$ &
$\displaystyle (-0.009174\pm 0.000161) + (5.370\pm 0.267)\cdot 10^{-6}i$ &

\\ 
$\displaystyle \lambda'_u $  &
$\displaystyle V_{ub}V_{us}^*$ &
$\displaystyle (0.000299\pm 0.000035) - (0.000760\pm 0.000027)i$ &

\\ 
$\displaystyle \lambda'_c $  &
$\displaystyle V_{cb}V_{cs}^*$ &
$\displaystyle (0.03944\pm 0.00069) + (1.249\pm 0.006)\cdot 10^{-6}i$ &

\\ \hline

\end{tabular}   $$

\label{tab:paramCKM}
\caption{\small \label{tab:paramCKM} CKM parameters.}
\end{table}
%%%%%%%%%%%%%%%%%%%%% END TABLE %%%%%%%%%%%%%%%%%%%%%%%%%%%%%
\vskip 10pt

\noindent The coefficients $A^h$ and $B^h$  in (\ref{4e},\ref{5e}) read for both modes and $h = 0, -$
\begin{equation}
\label{8e} 
A^0(\BsKK) = i {G_F \over \sqrt{2}}\ m_{B_s}^2 A_0^{\overline{B}_s \to \overline{K}^{*0}} (m_{K^{*0}}^2) f_{K^{*0}}
\end{equation}

\begin{equation}
\label{9e} 
A^-(\BsKK) = i {G_F \over \sqrt{2}}\ m_{B_s} m_{K^{*0}} F_-^{\overline{B}_s \to \overline{K}^{*0}} (m_{K^{*0}}^2) f_{K^{*0}}
\end{equation}

\begin{equation}
\label{10e} 
B^0(\BsKK) = B^-(\overline{B}_s \to K^{*0}\overline{K}^{*0}) = i {G_F \over \sqrt{2}}\ f_{B_s} f_{K^{*0}}^2
\end{equation}

\begin{equation}
\label{8bise} 
A^0(\BdKK) = i {G_F \over \sqrt{2}}\ m_{B_d}^2 A_0^{\overline{B}_d \to \overline{K}^{*0}} (m_{K^{*0}}^2) f_{K^{*0}}
\end{equation}

\begin{equation}
\label{9bise} 
A^-(\BdKK) = i {G_F \over \sqrt{2}}\ m_{B_d} m_{K^{*0}} F_-^{\overline{B}_d \to \overline{K}^{*0}} (m_{K^{*0}}^2) f_{K^{*0}}
\end{equation}

\begin{equation}
\label{10bise} 
B^0(\BdKK) = B^-(\overline{B}_d \to K^{*0}\overline{K}^{*0}) = i {G_F \over \sqrt{2}}\ f_{B_d} f_{K^{*0}}^2
\end{equation}

\vskip 10pt

\noindent where the relevant form factors and decay constants with their uncertainties are given in Table~\ref{tab:paramB} and Table~\ref{tab:param}.

\vskip 5 truemm

\begin{table}[htb]
\small
\centering
$$ \begin{tabular}{lccccccc}
\hline

$\displaystyle {\mathrm {particle}\ x} $ &
$\displaystyle {m_x\  (MeV)} $ & 
$\displaystyle {f_x (MeV)} $ & 
$\displaystyle {\tau_x (s)} $ & 
\\ 

\hline \hline

$\displaystyle \overline{B}_u$ &
$\displaystyle 5279.34 \pm 0$ &
$\displaystyle 190. \pm 5$ &
$\displaystyle 1.638 \cdot 10^{-12}\pm 0$ &

\\ 
$\displaystyle \overline{B}_d$ &
$\displaystyle 5279.65 \pm 0$ &
$\displaystyle 190. \pm 5$ &
$\displaystyle 1.519 \cdot 10^{-12}\pm 0$ &

\\ 
$\displaystyle \overline{B}_s$ &
$\displaystyle 5366.88 \pm 0$ &
$\displaystyle 230. \pm 5$ &
$\displaystyle 1.515 \cdot 10^{-12}\pm 0$ &

\\ \hline

\end{tabular}   $$

\label{tab:paramB}
\caption{\small \label{tab:paramB} $B$ meson parameters.}
\end{table}
%%%%%%%%%%%%%%%%%%%%% END TABLE %%%%%%%%%%%%%%%%%%%%%%%%%%%%%

%%%%%%%%%%%%%%%%%%%% Begin Table %%%%%%%%%%%%%%%%%%%%%%%%%%%%%%
\begin{table}[htb]
\small
\centering
$$ \begin{tabular}{lccccccc}
\hline

$\displaystyle {\mathrm {particle}\ x} $ &
$\displaystyle {m_x\  (MeV)} $ & 
$\displaystyle {f_x (MeV)} $ & 
$\displaystyle {A_0^{B_d\to x}} $ & 
$\displaystyle {F_-^{B_d\to x}} $ & 
$\displaystyle {A_0^{B_s\to x}} $ & 
$\displaystyle {F_-^{B_s\to x}} $ & 
\\

\hline \hline

$\displaystyle \rho^0$ &
$\displaystyle 775. \pm 0$ &
$\displaystyle 209 \pm 5$ &
$\displaystyle 0.30 \pm 0.05$ &
$\displaystyle 0.55 \pm 0.06$ &
$\displaystyle {\mathrm {n/a}}$ &
$\displaystyle {\mathrm {n/a}}$ &
\\ 
$\displaystyle \omega$ &
$\displaystyle 782. \pm 0$ &
$\displaystyle 187 \pm 5$ &
$\displaystyle 0.25 \pm 0.05$ &
$\displaystyle 0.50 \pm 0.06$ &
$\displaystyle {\mathrm {n/a}}$ &
$\displaystyle {\mathrm {n/a}}$ &
\\ 
$\displaystyle K^{*0}$ &
$\displaystyle 895. \pm 0$ &
$\displaystyle 218 \pm 5$ &
$\displaystyle 0.39 \pm 0.05$ &
$\displaystyle 0.68 \pm 0.06$ &
$\displaystyle 0.33 \pm 0.05$ &
$\displaystyle 0.53 \pm 0.06$ &
\\ 
$\displaystyle \phi$ &
$\displaystyle 1019.5 \pm 0$ &
$\displaystyle 221 \pm 5$ &
$\displaystyle {\mathrm {n/a}}$ &
$\displaystyle {\mathrm {n/a}}$ &
$\displaystyle 0.38 \pm 0.05$ &
$\displaystyle 0.65 \pm 0.06$ &
\\ \hline

\end{tabular}   $$

\label{tab:param}
\caption{\small \label{tab:param} Light meson parameters and heavy to light form factors.}
\end{table}
%%%%%%%%%%%%%%%%%%%%% END TABLE %%%%%%%%%%%%%%%%%%%%%%%%%%%%%
\vskip 10pt

\noindent In the particular case of $\BdsKK$, the rest of the relevant quantities in terms of the QCDF coefficients are given by the expressions \cite{BUCHALLA}
\begin{equation}
\label{11e} 
S^{p,h} = a_4^{p,h} - {1 \over 2}\ a_{10}^{p,h}
\end{equation}
\begin{equation}
\label{12e} 
T^h = b_3^h + 2b_4^h - {1 \over 2}\ b_3^{h,EW} - b_4^{h,EW}
\end{equation}

\noindent The coefficients $a_i^0, b_i^0$ for the helicity $h = 0$ are given in Tables~\ref{tab:a0_i} and~\ref{tab:b_i}, and $a_i^-, b_i^-$ for the transverse helicity $h = -$ in Tables~\ref{tab:aT_i} and~\ref{tab:b_i}. The main source of uncertainty is due to these QCDF coefficients, as made explicit in these Tables~\ref{tab:a0_i}-\ref{tab:b_i}. \par \noindent We assume that the amplitude for the transverse helicity $h = +$ is negligible, according to the hierarchy (\ref{eq:A=0},\ref{eq:F=F}).\par

%%%%%%%%%%%%%%%%%%%% Begin Table %%%%%%%%%%%%%%%%%%%%%%%%%%%%%%
\begin{table}[htb]
\small
\centering
$$ \begin{tabular}{lccccccc}
\hline

$\displaystyle {\mathrm {Coefficient} } $ &
$\displaystyle {\mathrm {Re}(a^0_i) } $ & 
$\displaystyle {\mathrm {Im}(a^0_i) } $ & 
$\displaystyle {\sigma_g(a^0_i) } $ & 

\\

\hline \hline

$\displaystyle a^0_1$ &
$\displaystyle  0.945\pm 1\%$ &
$\displaystyle 0.014 \pm 0\%$ &
$\displaystyle  \pm 10\%$ &
\\
$\displaystyle  a^0_2$ &
$\displaystyle 0.302 \pm 25\%$ &
$\displaystyle -0.081 \pm 0\%$ &
$\displaystyle \pm 10\%$ &
\\ 
$\displaystyle  a^0_3$ &
$\displaystyle -0.008 \pm 50\%$ &
$\displaystyle 0.003 \pm 0\%$ &
$\displaystyle \pm 10\%$ &
\\ 
$\displaystyle  a^{0u}_4$ &
$\displaystyle -0.021 \pm 7\%$ &
$\displaystyle -0.014 \pm 0\%$ &
$\displaystyle \pm 10\%$ &
\\ 
$\displaystyle  a^{0c}_4$ &
$\displaystyle -0.029\pm 5\%$ &
$\displaystyle -0.009 \pm 0\%$ &
$\displaystyle \pm 10\%$ &
\\ 
$\displaystyle  a^{0}_5$ &
$\displaystyle 0.015 \pm 33\%$ &
$\displaystyle -0.003 \pm 0\%$ &
$\displaystyle \pm 10\%$ &
\\
$\displaystyle  a^{0u}_7/\alpha$ &
$\displaystyle -0.271 \pm 110\%$ &
$\displaystyle -0.680 \pm 98\%$ &
$\displaystyle \pm 10\%$ &
\\ 
$\displaystyle  a^{0c}_7/\alpha$ &
$\displaystyle 0.020 \pm 20\%$ &
$\displaystyle 0.004 \pm 0\%$ &
$\displaystyle \pm 10\%$ &
\\
$\displaystyle  a^{0u}_9/\alpha$ &
$\displaystyle -1.365 \pm 22\%$ &
$\displaystyle -0.680 \pm 95\%$ &
$\displaystyle \pm 10\%$ &
\\ 
$\displaystyle  a^{0c}_9/\alpha$ &
$\displaystyle -1.058 \pm 2\%$ &
$\displaystyle -0.018 \pm 0\%$ &
$\displaystyle \pm 10\%$ &
\\
$\displaystyle  a^{0u}_{10}/\alpha$ &
$\displaystyle -0.334 \pm 25\%$ &
$\displaystyle 0.080 \pm 0\%$ &
$\displaystyle \pm 10\%$ &
\\ 
$\displaystyle  a^{0c}_{10}	/\alpha$ &
$\displaystyle -0.340 \pm 23\%$ &
$\displaystyle 0.083 \pm 0\%$ &
$\displaystyle \pm 10\%$ &

\\ \hline

\end{tabular}   $$

\label{tab:a0_i}
\caption{\small \label{tab:a0_i} The $a_i^0$ QCDF coefficients for the helicity $h=0$. $\sigma_g(a^0_i)$ is an additional error added in quadrature with the specific errors of $a^0_i$. }
\end{table}
%%%%%%%%%%%%%%%%%%%%% END TABLE %%%%%%%%%%%%%%%%%%%%%%%%%%%%%

%%%%%%%%%%%%%%%%%%%% Begin Table %%%%%%%%%%%%%%%%%%%%%%%%%%%%%%
\begin{table}[htb]
\small
\centering
$$ \begin{tabular}{lccccccc}
\hline

$\displaystyle {\mathrm {Coefficient} } $ &
$\displaystyle {\mathrm {Re}(a^-_i) } $ & 
$\displaystyle {\mathrm {Im}(a^-_i) } $ & 
$\displaystyle {\sigma_g(a^-_i) } $ & 

\\

\hline \hline

$\displaystyle a^-_1$ &
$\displaystyle  1.126\pm 1\%$ &
$\displaystyle 0.029 \pm 0\%$ &
$\displaystyle  \pm 10\%$ &
\\
$\displaystyle  a^-_2$ &
$\displaystyle -0.207 \pm 9\%$ &
$\displaystyle -0.162 \pm 0\%$ &
$\displaystyle \pm 10\%$ &
\\ 
$\displaystyle  a^-_3$ &
$\displaystyle 0.021 \pm 16\%$ &
$\displaystyle 0.005 \pm 0\%$ &
$\displaystyle \pm 10\%$ &
\\ 
$\displaystyle  a^{-u}_4$ &
$\displaystyle -0.045 \pm 1\%$ &
$\displaystyle -0.015 \pm 0\%$ &
$\displaystyle \pm 10\%$ &
\\ 
$\displaystyle  a^{-c}_4$ &
$\displaystyle -0.043\pm 1\%$ &
$\displaystyle -0.001 \pm 0\%$ &
$\displaystyle \pm 10\%$ &
\\ 
$\displaystyle  a^{-}_5$ &
$\displaystyle -0.026 \pm 15\%$ &
$\displaystyle -0.006 \pm 0\%$ &
$\displaystyle \pm 10\%$ &
\\
$\displaystyle  a^{-u}_7/\alpha$ &
$\displaystyle 1.052 \pm 25\%$ &
$\displaystyle 0.009 \pm 0\%$ &
$\displaystyle \pm 10\%$ &
\\ 
$\displaystyle  a^{-c}_7/\alpha$ &
$\displaystyle 1.024 \pm 25\%$ &
$\displaystyle 0.009 \pm 0\%$ &
$\displaystyle \pm 10\%$ &
\\
$\displaystyle  a^{-u}_9/\alpha$ &
$\displaystyle -0.279 \pm 99\%$ &
$\displaystyle -0.037 \pm 0\%$ &
$\displaystyle \pm 10\%$ &
\\ 
$\displaystyle  a^{-c}_9/\alpha$ &
$\displaystyle -0.307 \pm 90\%$ &
$\displaystyle -0.037 \pm 0\%$ &
$\displaystyle \pm 10\%$ &
\\
$\displaystyle  a^{-u}_{10}/\alpha$ &
$\displaystyle 0.292 \pm 21\%$ &
$\displaystyle 0.171 \pm 0\%$ &
$\displaystyle \pm 10\%$ &
\\ 
$\displaystyle  a^{-c}_{10}	/\alpha$ &
$\displaystyle 0.293 \pm 21\%$ &
$\displaystyle 0.182 \pm 0\%$ &
$\displaystyle \pm 10\%$ &

\\ \hline

\end{tabular}   $$

\label{tab:aT_i}
\caption{\small \label{tab:aT_i} The $a_i^-$ QCDF coefficients for the transverse helicity $h=-$. $\sigma_g(a^-_i)$ is an additional error added in quadrature with the specific errors of $a^-_i$.}
\end{table}
%%%%%%%%%%%%%%%%%%%%% END TABLE %%%%%%%%%%%%%%%%%%%%%%%%%%%%%

%%%%%%%%%%%%%%%%%%%% Begin Table %%%%%%%%%%%%%%%%%%%%%%%%%%%%%%
\begin{table}[htb]
\small
\centering
$$ \begin{tabular}{lccccccc}
\hline

$\displaystyle {\mathrm {index}\ i} $ &
$\displaystyle {\mathrm {Re}(b^0_i) } $ & 
$\displaystyle {\mathrm {Im}(b^0_i) } $ & 
$\displaystyle {\mathrm {Re}(b^-_i) } $ & 
$\displaystyle {\mathrm {Im}(b^-_i) } $ & 

\\

\hline \hline

$\displaystyle 1$ &
$\displaystyle  9.692\pm 1.110$ &
$\displaystyle -4.052 \pm 1.060$ &
$\displaystyle  0.691\pm 0.040$ &
$\displaystyle 0\pm 0 $ &
\\
$\displaystyle  2$ &
$\displaystyle -3.038 \pm0.347$ &
$\displaystyle 1.268 \pm 0.331$ &
$\displaystyle  -0.0217\pm 0.013$ &
$\displaystyle 0\pm 0 $ &

\\ 
$\displaystyle  3$ &
$\displaystyle 3.372 \pm 0.915$ &
$\displaystyle -3.784 \pm 1.210$ &
$\displaystyle  -3.736\pm 0.929$ &
$\displaystyle 3.855\pm 1.210 $ &

\\ 
$\displaystyle  4$ &
$\displaystyle -1.203 \pm 0.138$ &
$\displaystyle 0.503 \pm 0.131$ &
$\displaystyle  -0.086\pm 0.005$ &
$\displaystyle 0\pm 0 $ &

\\ 
$\displaystyle  3_{EW}$ &
$\displaystyle -0.123\pm 0.012$ &
$\displaystyle0.080 \pm 0.023$ &
$\displaystyle  0.0305\pm 0.0094$ &
$\displaystyle -0.0399\pm 0.0126 $ &

\\ 
$\displaystyle  4_{EW}$ &
$\displaystyle 0.035 \pm 0.004$ &
$\displaystyle -0.015 \pm 0.004$ &
$\displaystyle  0.0025\pm 0.0001$ &
$\displaystyle 0\pm 0 $ &

\\ \hline

\end{tabular}   $$

\label{tab:b_i}
\caption{\small \label{tab:b_i} The $b_i^0$ and $b_i^-$ QCDF annihilation coefficients for the helicity $h=0$ and $h=-$. }
\end{table}
%%%%%%%%%%%%%%%%%%%%% END TABLE %%%%%%%%%%%%%%%%%%%%%%%%%%%%%

\vskip 10pt

\noindent In the present paper we only make explicit the calculations and results for the decay modes $\BdsKK$. The rest of the decay modes for which there are presently experimental data, made explicit in Table~\ref{tab:B_decays} above, will be studied in detail in a forthcoming paper \cite{AO:2}, and we only quote here the final results. Gathering all the parameters we find the expected values of the different observables, BR, $f_L$ and $A_{CP}$ obtained within QCD Factorization, as made explicit in Table~\ref{tab:B_decays_QCDF}. The various contributing errors for each parameter are assumed to be a flat distribution in the uncertainty range given in the Tables~\ref{tab:paramCKM} to~\ref{tab:b_i}.
%%%%%%%%%%%%%%%%%%%% Figure Feynman diagram Bs to DsK  %%%%%%%%%%%%%%%%%%%
\begin{figure}[hbt]
\vfill
\begin{center}

%\vskip 5cm
\includegraphics[width=1.\textwidth]{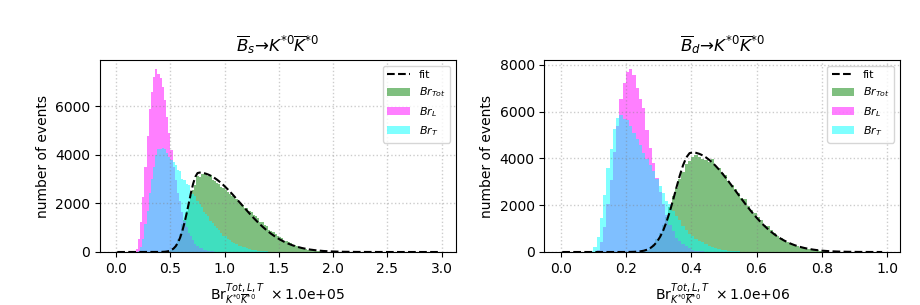}

\caption{\small \label{fig:Br} The CP-averaged Branching Fractions for a selection of ${\rm \overline{B}}_{d,s}$ decays to $\rm K^{*0}\overline{K}^{*0}$  final states. The written values correspond to QCD Factorization.}
\end{center}
\vfill

\end{figure}
%%%%%%%%%%%%%%%%%%%%%%%%% End Figure %%%%%%%%%%%%%%%%%%%%%%%%%
%%%%%%%%%%%%%%%%%%%% Figure Feynman diagram Bs to DsK  %%%%%%%%%%%%%%%%%%%
\begin{figure}[hbt]
\vfill
\begin{center}

%\vskip 5cm
\includegraphics[width=1.\textwidth]{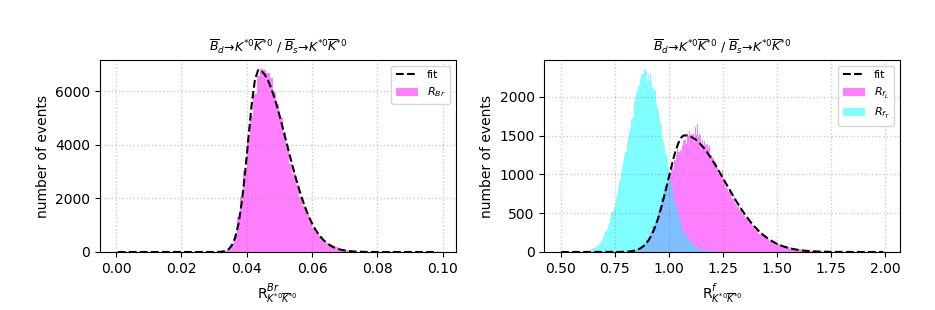}

\caption{\small \label{fig:r_all} The ratio ${\rm Br}({\BdKK})/{\rm Br}({\BsKK})$ and $f_L({\BdKK })/f_L({\BsKK})$ as calculated with QCD Factorization.}
\end{center}
\vfill

\end{figure}
%%%%%%%%%%%%%%%%%%%%%%%%% End Figure %%%%%%%%%%%%%%%%%%%%%%%%%

%\begin{adjustwidth}{-1cm}{-1cm}
%\newgeometry{hmargin={1cm}}
   
\begin{table}[htb]
\centering
\small
$$  \hspace{-2cm}
\begin{tabular}{lcccccc}

$\displaystyle {\mathrm {{B}\ decay}} $ &
$\displaystyle {\mathrm {Br} (\times10^{-6})} $ & 
$\displaystyle {f_L} $ &
$\displaystyle {f_\parallel} $  &
$\displaystyle {f_\perp} $ &
$\displaystyle {A_{\rm CP}} $ \\
\hline \hline

\textcolor{red} {\underline{ {\boldmath $\overline{B}_s $} modes}} \\ 

$\displaystyle \overline{B}_s\to \phi\rho^0$ &
$\displaystyle {\mathrm {0.306\pm0.059}}$ &
$\displaystyle {\mathrm {0.951\pm0.023}} $ &
$\displaystyle {\mathrm {{\it 0.025\pm0.012}}} $ &
$\displaystyle {\mathrm {{\it 0.025\pm0.012}}} $ &
$\displaystyle {\mathrm {0.318\pm 0.034}} $ \\

$\displaystyle \overline{B}_s\to \phi\phi$ &
$\displaystyle {\mathrm {22.630\pm6.670}}$ &
$\displaystyle {\mathrm {0.315 \pm 0.070}} $ &
$\displaystyle {\mathrm {{\it 0.343\pm 0.035}}} $ &
$\displaystyle {\mathrm {{\it 0.343\pm 0.035}}} $ &
$\displaystyle {\mathrm {0.007 \pm 0.002}} $ 
\\

$\displaystyle \overline{B}_s\to \overline{K}^{*0}{K}^{*0}$ &
$\displaystyle {\mathrm {10.050\pm2.640}}$ &
$\displaystyle {\mathrm {0.429 \pm 0.088}} $ &
$\displaystyle {\mathrm {{\it 0.286\pm 0.044}}} $ &
$\displaystyle {\mathrm {{\it 0.286\pm 0.044}}} $ & 
$\displaystyle {\mathrm {0.006\pm 0.001}} $ \\

$\displaystyle \overline{B}_s\to \phi{K}^{*0}$ &
$\displaystyle {\mathrm {0.504\pm0.163}}$ &
$\displaystyle {\mathrm {0.365 \pm 0.074}} $ &
$\displaystyle {\mathrm {{\it 0.318\pm 0.037 }}} $ &
$\displaystyle {\mathrm {{\it 0.318\pm 0.037 }}} $ &
$\displaystyle {\mathrm {-0.160\pm 0.037}} $ \\

\textcolor{red} {\underline{ {\boldmath $\overline{B}_d $} modes}}\\ 

$\displaystyle \overline{B}^0\to \omega \omega$ &
$\displaystyle {\mathrm {0.840\pm0.204}}$ &
$\displaystyle {\mathrm {0.914\pm0.024}} $ &
$\displaystyle {\mathrm {{\it 0.043\pm 0.012 }}} $ &
$\displaystyle {\mathrm {{\it 0.043\pm 0.012 }}} $ &
$\displaystyle {\mathrm {-0.416\pm0.071}} $ \\

$\displaystyle \overline{B}^0\to \rho^{+}\rho^{-}$ &
$\displaystyle {\mathrm {23.840\pm4.910}}$ &
$\displaystyle {\mathrm {0.904 \pm 0.023}} $ &
$\displaystyle {\mathrm {{\it 0.048\pm 0.012 }}} $ &
$\displaystyle {\mathrm {{\it 0.048\pm 0.012 }}} $ &
$\displaystyle {\mathrm {-0.071\pm 0.015}} $ \\

$\displaystyle \overline{B}^0\to \rho^{0}\rho^{0}$ &
$\displaystyle {\mathrm {0.734\pm 0.203}}$ &
$\displaystyle {\mathrm {0.841 \pm  0.061}} $ &
$\displaystyle {\mathrm {{\it 0.080\pm 0.031 }}} $ &
$\displaystyle {\mathrm {{\it 0.080\pm 0.031 }}} $ &
$\displaystyle {\mathrm {0.572\pm 0.099}} $ \\

$\displaystyle \overline{B}^0\to \rho^+ K^{*-}$ &
$\displaystyle {\mathrm {6.973\pm2.000}}$ &
$\displaystyle {\mathrm {0.405 \pm 0.062}} $ &
$\displaystyle {\mathrm {{\it 0.298\pm 0.031 }}} $ &
$\displaystyle {\mathrm {{\it 0.298\pm 0.031 }}} $ &
$\displaystyle {\mathrm {0.305\pm  0.065}} $ \\

$\displaystyle \overline{B}^0\to \rho^0 \overline{K}^{*0}$ &
$\displaystyle {\mathrm {3.446\pm1.110}}$ &
$\displaystyle {\mathrm {0.324 \pm 0.070}} $ &
$\displaystyle {\mathrm {{\it 0.338\pm 0.035}}} $ &
$\displaystyle {\mathrm {{\it 0.338\pm 0.035}}} $ &
$\displaystyle {\mathrm {-0.189 \pm {0.040}}} $\\

$\displaystyle \overline{B}^0\to \omega \overline{K}^{*0}$ &
$\displaystyle {\mathrm {2.932\pm0.951}}$ &
$\displaystyle {\mathrm {0.387 \pm 0.085}} $ &
$\displaystyle {\mathrm {{\it 0.307\pm 0.043}}} $ &
$\displaystyle {\mathrm {{\it 0.307\pm 0.043}}} $ &
$\displaystyle {\mathrm {0.187 \pm {0.047}}} $ 
\\

$\displaystyle \overline{B}^0\to \overline{K}^{*0}\phi$ &
$\displaystyle {\mathrm {9.285\pm 2.550}}$ &
$\displaystyle {\mathrm {0.340 \pm 0.071}} $ &
$\displaystyle {\mathrm {{\it 0.330\pm 0.036}}} $ &
$\displaystyle {\mathrm {{\it 0.330\pm 0.036}}} $ &
$\displaystyle {\mathrm {0.009\pm 0.002}} $ \\

$\displaystyle \overline{B}^0\to K^{*0}\overline{K}^{*0}$ &
$\displaystyle {\mathrm {0.465\pm0.095}}$ &
$\displaystyle {\mathrm {0.498 \pm 0.086}} $ &
$\displaystyle {\mathrm {{\it 0.251\pm 0.043}}} $ &
$\displaystyle {\mathrm {{\it 0.251\pm 0.043}}} $ &
$\displaystyle {\mathrm {-0.165\pm 0.022}} $ \\ 
\hline \hline

\textcolor{red} {\underline{ {\boldmath $\overline{B}_u $} modes}}\\ 

$\displaystyle B^-\to \omega {K}^{*-}$ &
$\displaystyle {\mathrm {3.425\pm 0.990 }}$ &
$\displaystyle {\mathrm {0.419 \pm 0.076}} $ &
$\displaystyle {\mathrm {{\it 0.291\pm 0.038}}} $ &
$\displaystyle {\mathrm {{\it 0.291\pm 0.038}}} $ &
$\displaystyle {\mathrm {0.408\pm 0.087}} $ \\

$\displaystyle B^-\to \omega{\rho}^{-}$ &
$\displaystyle {\mathrm {13.140\pm2.720}}$ &
$\displaystyle {\mathrm {0.924 \pm 0.022}} $ &
$\displaystyle {\mathrm {{\it 0.038\pm 0.011}}} $ &
$\displaystyle {\mathrm {{\it 0.038\pm 0.011}}} $ &
$\displaystyle {\mathrm {-0.189\pm 0.037}} $ \\

$\displaystyle B^-\to \rho^0{\rho}^{-}$ &
$\displaystyle {\mathrm {17.920\pm3.910}}$ &
$\displaystyle {\mathrm {0.956 \pm 0.012}} $ &
$\displaystyle {\mathrm {{\it 0.022\pm 0.006}}} $ &
$\displaystyle {\mathrm {{\it 0.022\pm 0.006}}} $ &
$\displaystyle {\mathrm {-0.000\pm 0.001}} $ \\

$\displaystyle B^-\to \rho^0{K}^{*-}$ &
$\displaystyle {\mathrm {4.559\pm1.000}}$ &
$\displaystyle {\mathrm {0.535 \pm 0.093}} $ &
$\displaystyle {\mathrm {{\it 0.233\pm 0.0047}}} $ &
$\displaystyle {\mathrm {{\it 0.233\pm 0.0047}}} $ &
$\displaystyle {\mathrm {0.435\pm 0.054}} $ \\

$\displaystyle B^-\to \rho^-\overline{K}^{*0}$ &
$\displaystyle {\mathrm {7.242\pm2.070}}$ &
$\displaystyle {\mathrm {0.413 \pm 0.082}} $ &
$\displaystyle {\mathrm {{\it 0.294\pm 0.0041}}} $ &
$\displaystyle {\mathrm {{\it 0.294\pm 0.0041}}} $ &
$\displaystyle {\mathrm {-0.000\pm 0.001}} $ \\

$\displaystyle B^-\to K^{*-}\phi $ &
$\displaystyle {\mathrm {9.803\pm2.630}}$ &
$\displaystyle {\mathrm {0.337 \pm 0.071}} $ &
$\displaystyle {\mathrm {{\it 0.332\pm 0.036}}} $ &
$\displaystyle {\mathrm {{\it 0.332\pm 0.036}}} $ &
$\displaystyle {\mathrm {0.003\pm 0.034}} $ \\

$\displaystyle B^-\to K^{*-}{K}^{*0}$ &
$\displaystyle {\mathrm {0.443\pm 0.106}}$ &
$\displaystyle {\mathrm {0.464 \pm 0.081}} $ &
$\displaystyle {\mathrm {{\it 0.268\pm 0.041}}} $ &
$\displaystyle {\mathrm {{\it 0.268\pm 0.041}}} $ &
$\displaystyle {\mathrm {-0.027\pm 0.034}} $ \\
	
\hline\hline

\hline

\end{tabular}   $$

\label{tab:B_decays_QCDF}
\caption{\small \label{tab:B_decays_QCDF} B-meson branching fractions, $f_L$, $f_\parallel$, $f_\perp$ and $A_{CP}$  for some selected V$_1$-V$_2$ modes as calculated with QCD Factorization. The values in italics are derived from $f_\parallel = f_\perp = f_T/2= (1-f_L)/2$. This table is to be compared with the experimental Table~\ref{tab:B_decays}.}

\end{table} 
%\end{adjustwidth}
%\restoregeometry
%%%%%%%%%%%%%%%%%%%% Figure Feynman diagram Bs to DsK  %%%%%%%%%%%%%%%%%%%
\begin{figure}[bt]
\vfill
\begin{center}

%\vskip 5cm
\includegraphics[width=1.\textwidth]{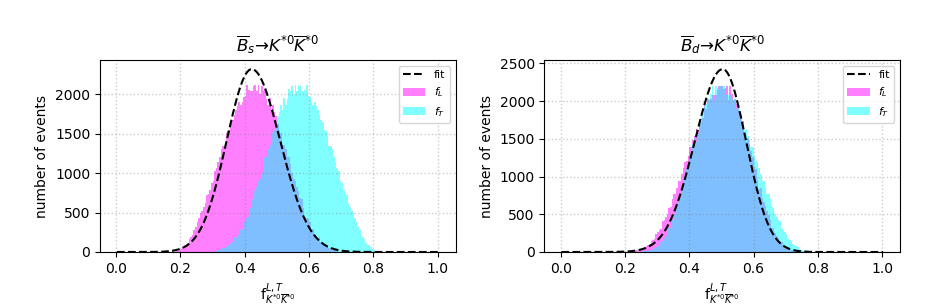}

\caption{\small \label{fig:f_all} The CP-averaged fractions of longitudinal and transverse polarizations ($f_{L}, f_T$) for the decays of $\BdsKK$ as calculated with QCD Factorization. }
\end{center}
\vfill

\end{figure}
%%%%%%%%%%%%%%%%%%%%%%%%% End Figure %%%%%%%%%%%%%%%%%%%%%%%%%

\noindent From the values for the form factors, decay constants and QCDF coefficients of the Tables~\ref{tab:paramCKM}-\ref{tab:b_i}, one finds the values of the branching ratios (see Figure~\ref{fig:Br})
\begin{equation}
\begin{array}{cccc} 
{\rm Br }(\BdKK) & = & (0.40^{+0.14}_{-0.05})\times 10^{-6} \\
\\
{\rm Br }(\BsKK) & = & (7.68^{+0.38}_{-0.10})\times 10^{-6} \\
\end{array}
\label{14e} 
\end{equation}

\noindent that give the ratio (see Figure~\ref{fig:r_all})
\begin{equation} 
\left[{{\rm Br}(\BdKK) \over {\rm Br}(\BsKK)}\right]_{\rm QCDF} = 0.044^{+0.008}_{-0.003}
\label{15e} 
\end{equation}

\noindent that is very close to the naive value (\ref{2e}), i.e. with no evidence of $U$-spin breaking.\par
\noindent For the longitudinal fractions one finds, on the other hand, the central values (see Figure~\ref{fig:f_all})
\begin{equation}
\begin{array}{cccc} 
f_L(\BdKK) & = & 0.50^{+0.08}_{-0.08} \\
\\
f_L(\BsKK) & = & 0.42^{+0.09}_{-0.08} \\
\end{array}
\label{16e} 
\end{equation}
\noindent that give the ratio (see Figure~\ref{fig:r_all})
\begin{equation}
\label{17e} 
\left[{f_L(\BdKK) \over f_L(\BsKK)}\right]_{\rm QCDF} = 1.07^{+0.19}_{-0.08}
\end{equation}

\noindent with some small evidence of a theoretical U-spin breaking.\par 
\vskip 10pt
\noindent The values (\ref{16e}) are nevertheless close to each other and are at serious odds with experiment~(\ref{eq:LHCb_Bs}) and~(\ref{eq:LHCb_Bd}) for $f_L$,  and neither of these values agrees with experiment. The quantitative comparison of LHCb data in equation~(\ref{eq:f_Lratio}) with QCDF leads to
\begin{equation}
\label{18e} 
\left[{f_L(\BdKK) \over f_L(\BsKK)}\right]_{\rm LHCb}  -
\left[{f_L(\BdKK) \over f_L(\BsKK)}\right]_{\rm QCDF} = 1.96^{+0.56}_{-0.60}
\end{equation}
\noindent which is a 3.3 standard deviation (std) effect. Should the experimental error on $f_L(\BdKK) / f_L(\BsKK)$ be reduced by a factor 2, the significance would exceed 5 std. Furthermore the expectation $f_\parallel\simeq f_\perp$ is strongly contradicted by the experimental LHCb data quoted in~(\ref{eq:LHCb_Bs}).

\vskip 10pt
\noindent To compare the QCD Factorization predictions to the data, one should consider the Form Factor $F_+ \neq 0$, which will lead to $f_\parallel \neq f_\perp$. Equation~(\ref{eq:FF}) is used with values of $A_1$ and $V$ estimated using light-cone sum rules~\cite{BZ:1}. As expected the central values $F_+^{B\to V}\simeq 0$ is obtained, though with some uncertainties. We follow the estimates from~\cite{BRY:1} and use the values in Table~\ref{tab:Fplus}.
%%%%%%%%%%%%%%%%%%%% Figure Feynman diagram Bs to DsK  %%%%%%%%%%%%%%%%%%%
\begin{figure}[hbt]
\vfill
\begin{center}

%\vskip 5cm
\includegraphics[width=0.9\textwidth]{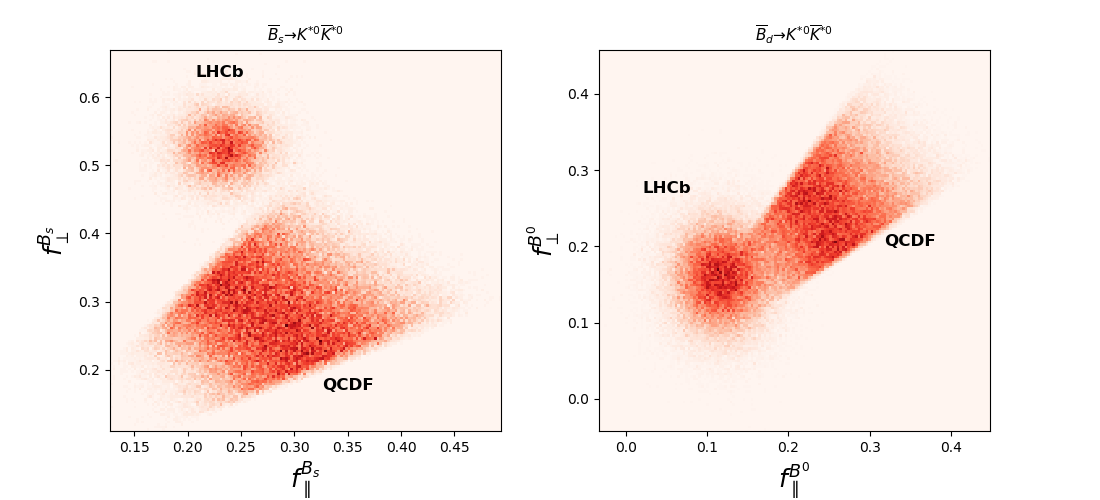}

\caption{\small \label{fig:fperp_vs_fpara} $f_\perp$ versus $f_\parallel$ as predicted by QCD Factorization using the form factors $F_+$ and $F_-$  obtained from light-cone sum rules. The data from LHCb~\cite{lhcb:1} are also displayed.}
\end{center}
\vfill

\end{figure}
%%%%%%%%%%%%%%%%%%%%%%%%% End Figure %%%%%%%%%%%%%%%%%%%%%%%%%
%%%%%%%%%%%%%%%%%%%% Begin Table %%%%%%%%%%%%%%%%%%%%%%%%%%%%%%
\begin{table}[htb]

\centering
\small 
$$ \begin{tabular}{lccccccc}
\hline

$\displaystyle {\mathrm {particle }\ x } $ &
$\displaystyle {\rho} $ & 
$\displaystyle {\omega} $ & 
$\displaystyle {K^\ast} $ & 
$\displaystyle {\phi} $ 
\\

\hline \hline

$\displaystyle F_+^{B_d\to x}$ &
$\displaystyle 0. \pm 0.06$ &
$\displaystyle 0. \pm 0.06$ &
$\displaystyle 0. \pm 0.06$ &
$\displaystyle {\mathrm {n/a}}$ &
\\ 
$\displaystyle F_+^{B_s\to x}$ &
$\displaystyle {\mathrm {n/a}}$ &
$\displaystyle {\mathrm {n/a}}$ &
$\displaystyle 0. \pm 0.06$ &
$\displaystyle 0. \pm 0.06$ &

\\ \hline

\end{tabular}   $$

\label{tab:Fplus}
\caption{\small \label{tab:Fplus} $F_+$ form factors.}
\end{table}
%%%%%%%%%%%%%%%%%%%%% END TABLE %%%%%%%%%%%%%%%%%%%%%%%%%%%%%

\noindent We summarize in Figure~\ref{fig:fperp_vs_fpara} the prediction from QCD Factorization, using the values of $F_+$ as given in Table~\ref{tab:Fplus}. We conclude that the expectation of QCD Factorization is not compatible with the data for the decay ${\rm \overline{B}}_{s}\to \rm K^{*0}\overline{K}^{*0}$  and it is marginally compatible for ${\rm \overline{B}}_{d}\to \rm K^{*0}\overline{K}^{*0}$.

% ================================ new section ===============================

\section{Sensitivities at FCC-ee}

We estimate now the sensitivities that can be attained at FCC-ee. To this end we have simulated a generic detector with parametrized resolutions described in detail in~\cite{AOP:1}. 
\subsection { Detector simulation}
\noindent In the following we summarize the main characteristics of the simulated detector. The components of this detector, which are relevant for this study, include a silicon pixelized vertex system, a large gaseous tracking device, an outer silicon wrapper and a ToF detector embedded in a soleno\"{\i}d with a field of 2 Tesla. The parametrizations of the vertex and tracking resolutions are obtained by the simulation of the following components :
\begin{itemize}
\item {\it Silicon vertex and inner tracker : }
The silicon detector includes 8 layers of pixel sensors with the main parameters listed in Table~\ref{tab:vertex}. It also includes 6 layers of disks in each end cap (Table~\ref{tab:disks}).
%%%%%%%%%%%%%%%%%%%% Begin Table %%%%%%%%%%%%%%%%%%%%%%%%%%%%%%
\begin{table}[htb]

\centering
\small
$$ \begin{tabular}{lccccccccc}
\hline

$\displaystyle {layer} $ &
$\displaystyle {1} $ & 
$\displaystyle {2} $ & 
$\displaystyle {3} $ & 
$\displaystyle {4} $ &
$\displaystyle {5} $ & 
$\displaystyle {6} $ & 
$\displaystyle {7} $ & 
$\displaystyle {8} $ 
\\

\hline \hline

$\displaystyle r (cm)$ &
$\displaystyle 1.2$ &
$\displaystyle 1.8$ &
$\displaystyle 3.7$ &
$\displaystyle 6.0$ &
$\displaystyle 10.0$ &
$\displaystyle 20.0$ &
$\displaystyle 30.0$ &
$\displaystyle 38.0$ &
\\
$\displaystyle z (\pm cm)$ &
$\displaystyle 20.0$ &
$\displaystyle 23.0$ &
$\displaystyle20.0$ &
$\displaystyle 60.0$ &
$\displaystyle 60.0$ &
$\displaystyle 60.0$ &
$\displaystyle 60.0$ &
$\displaystyle 64.0$ &
\\
$\displaystyle X/X_0$ &
$\displaystyle {\mathrm {0.001}}$ &
$\displaystyle {\mathrm {0.001}}$ &
$\displaystyle {\mathrm {0.001}}$ &
$\displaystyle {\mathrm {0.001}}$ &
$\displaystyle {\mathrm {0.001}}$ &
$\displaystyle {\mathrm {0.001}}$ &
$\displaystyle {\mathrm {0.007}}$ &
$\displaystyle {\mathrm {0.007}}$ &
\\

$\displaystyle \sigma (\mu m)$ &
$\displaystyle {\mathrm {3.0}}$ &
$\displaystyle {\mathrm {3.0}}$ &
$\displaystyle {\mathrm {4.0}}$ &
$\displaystyle {\mathrm {4.0}}$ &
$\displaystyle {\mathrm {4.0}}$ &
$\displaystyle {\mathrm {4.0}}$ &
$\displaystyle {\mathrm {7.0}}$ &
$\displaystyle {\mathrm {7.0}}$ &

\\ \hline

\end{tabular}   $$

\label{tab:vertex}
\caption{\small \label{tab:vertex} Barrel silicon vertex and inner tracker parameters.}
\end{table}
%%%%%%%%%%%%%%%%%%%%% END TABLE %%%%%%%%%%%%%%%%%%%%%%%%%%%%%
%%%%%%%%%%%%%%%%%%%% Begin Table %%%%%%%%%%%%%%%%%%%%%%%%%%%%%%
\begin{table}[htb]

\centering
\small
$$ \begin{tabular}{lccccccccc}
\hline

$\displaystyle {layer} $ &
$\displaystyle {1} $ & 
$\displaystyle {2} $ & 
$\displaystyle {3} $ & 
$\displaystyle {4} $ &
$\displaystyle {5} $ & 
$\displaystyle {6} $ 
\\

\hline \hline

$\displaystyle {\rm inner}R (cm)$ &
$\displaystyle 5.5$ &
$\displaystyle 5.5$ &
$\displaystyle 11.5$ &
$\displaystyle 11.5$ &
$\displaystyle 16.5$ &
$\displaystyle16.5$ &

\\
$\displaystyle {\rm outer}R (cm)$ &
$\displaystyle 38.5$ &
$\displaystyle 38.5$ &
$\displaystyle 38.5$ &
$\displaystyle 38.5$ &
$\displaystyle 38.5$ &
$\displaystyle 38.5$ &

\\
$\displaystyle z ( cm)$ &
$\displaystyle 65$ &
$\displaystyle 80$ &
$\displaystyle150$ &
$\displaystyle 170$ &
$\displaystyle 220$ &
$\displaystyle 240$ &

\\
$\displaystyle X/X_0$ &
$\displaystyle {\mathrm {0.003}}$ &
$\displaystyle {\mathrm {0.003}}$ &
$\displaystyle {\mathrm {0.003}}$ &
$\displaystyle {\mathrm {0.003}}$ &
$\displaystyle {\mathrm {0.003}}$ &
$\displaystyle {\mathrm {0.003}}$ &

\\

$\displaystyle \sigma (\mu m)$ &
$\displaystyle {\mathrm {7.0}}$ &
$\displaystyle {\mathrm {7.0}}$ &
$\displaystyle {\mathrm {7.0}}$ &
$\displaystyle {\mathrm {7.0}}$ &
$\displaystyle {\mathrm {7.0}}$ &
$\displaystyle {\mathrm {7.0}}$ &

\\ \hline

\end{tabular}   $$

\label{tab:disks}
\caption{\small \label{tab:disks}End cap silicon vertex and inner tracker parameters.}
\end{table}
%%%%%%%%%%%%%%%%%%%%% END TABLE %%%%%%%%%%%%%%%%%%%%%%%%%%%%%

\item {\it TPC :} The central tracking is achieved by a TPC, the main parameters of which are listed in Table~\ref{tab:TPC}.
%%%%%%%%%%%%%%%%%%%% Begin Table %%%%%%%%%%%%%%%%%%%%%%%%%%%%%%
\begin{table}[htb]

\centering
\small 
$$ \begin{tabular}{lccccccccc}
\hline

$\displaystyle {\rm inner R} $ &
$\displaystyle {\rm outer R} $ & 
$\displaystyle {z} $ & 
$\displaystyle {\rm Number} $ &
$\displaystyle {\sigma(r)} $ & 
$\displaystyle {\sigma(z)} $ 
\\
$\displaystyle {(cm)} $ &
$\displaystyle {(cm)} $ & 
$\displaystyle {(cm)} $ & 
$\displaystyle {\rm layers} $ &
$\displaystyle {\mu m} $ & 
$\displaystyle {\mu m} $ 
\\

\hline \hline

$\displaystyle 40$ &
$\displaystyle 220$ &
$\displaystyle \pm 250$ &
$\displaystyle 250$ &
$\displaystyle 100$ &
$\displaystyle 500$ &
\\ \hline

\end{tabular}   $$

\label{tab:TPC}
\caption{\small \label{tab:TPC} The TPC tracker parameters.}
\end{table}
%%%%%%%%%%%%%%%%%%%%% END TABLE %%%%%%%%%%%%%%%%%%%%%%%%%%%%%
\item {\it Outer wrapper :}
A silicon outer wrapper  surrounding the TPC is included with a point resolution of $10\mu m$.
\end{itemize}
The particle identification (PID) is achieved using a dedicated time-of-flight ($ToF$) system and the cluster counting ($dN/dx$) from the gaseous tracker with the resolutions $\sigma(ToF) = 10ps$ and $\sigma(dN/dx) =2.2\%$, respectively. With the tracking parameters above, some important resolutions are obtained for the decay $\BdsKK$.  It includes in particular the reconstructed error on B mass and on the B flight distance:
\begin{equation}
\begin{array}{cccc} 
\sigma_{m_B} & \simeq & 6\ MeV \\
\sigma_{d_B} & \simeq & 20\ \mu m \\
\end{array}
\label{19e} 
\end{equation}

\begin{figure}[hbt]
\vfill
\begin{center}

%\vskip 5cm
\includegraphics[width=0.49\textwidth]{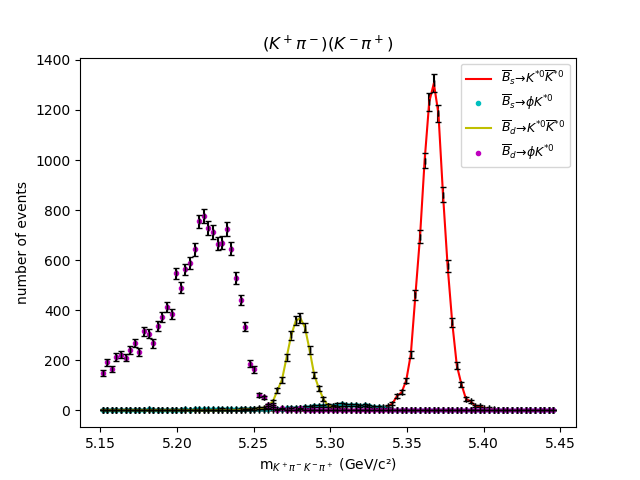}
\includegraphics[width=0.49\textwidth]{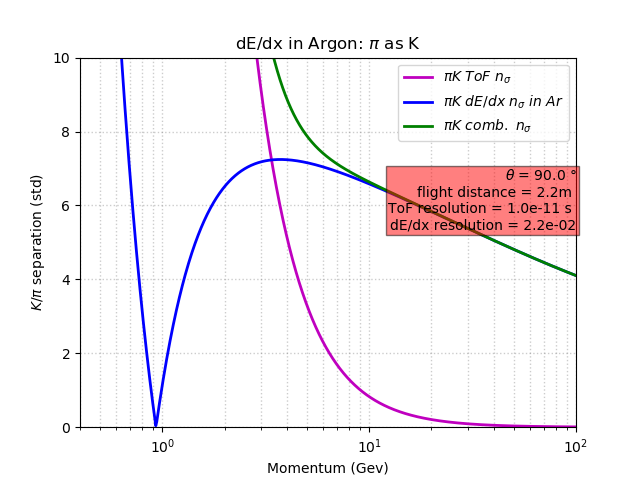}

\caption{\label{fig:resolutions} Invariant mass for ${\rm \overline{B}}_{d,s}\to  \rm (K^+\pi^-)(K^-\pi^+)$ combinations without particle identification (left) and $K/\pi$ separation expected with the simulated detector (right). The left plot shows also the contributions from ${\rm \overline{B}}_{d}\to \rm \phi \overline{K}^{*0}$ and ${\rm \overline{B}}_{s}\to  \rm \phi K^{*0}$ where one of the K from $\phi$ is misidentify as a $\pi$. Combinatoric background is not shown neither possible contributions coming from $\rm K^*_0(700)$ , $\rm K^*_0(1430)$ ..., which are however low.}
\end{center}

\end{figure}

\noindent These figures are instrumental to suppress the background very efficiently.  Figure~\ref{fig:resolutions} shows the $({\rm K}^+\pi^-)({\rm K}^-\pi^+)$ mass for the channel ${\rm \overline{B}}_{s}\to \rm K^{*0}\overline{K}^{*0}$ and the $\rm K/\pi$ separation. It should also be stressed that PID is very important for carrying out this study. Indeed, the combinatoric background from ${\rm Z}\to q\overline{q}$ (mostly ${\rm Z}\to b\overline{b}$) as well as  final states such as $\rm \overline{B} \to (K^+K^-)_{\phi}(K^-\pi^+)_{\overline{K}^{*0}}$ may lead to significant contribution if one has no or poor $\rm K/\pi$ separation.

%%%%%%%%%%%%%%%%%%%% Figure Feynman diagram Bs to DsK  %%%%%%%%%%%%%%%%%%%
%%%%%%%%%%%%%%%%%%%%%%%%% End Figure %%%%%%%%%%%%%%%%%%%%%%%%%

\begin{table}[btb]
\small
\centering
$$ \begin{tabular}{cccc}
\hline
&  & $\displaystyle {\mathrm {E_{cm} = m_Z \ and\  \int L = 150 ab^{-1}}}$   &  \\

$\displaystyle {\mathrm {\sigma (e^+e^- \to Z )}} $ &
$\displaystyle {\mathrm {number}} $ &
$\displaystyle {\mathrm {f(Z\to \overline{B_s})}} $ &
$\displaystyle {\mathrm {Number \ of}} $ \\

$\displaystyle {\mathrm {nb}} $ & 
$\displaystyle {\mathrm {of \ Z} } $ &
$\displaystyle {\mathrm{}} $ &
$\displaystyle {\mathrm{produced \ \overline{B} }} $\\ 
\hline \hline 

$\displaystyle \sim 42.9$ &
$\displaystyle {\mathrm {\sim 6.4\ 10^{12}}}$ &
$\displaystyle {\mathrm {0.0159}} $ &
$\displaystyle \sim 1\ 10^{11} \overline{\rm B}_s $\\ 
$\displaystyle \sim 42.9$ &
$\displaystyle {\mathrm {\sim 6.4\ 10^{12}}}$ &
$\displaystyle {\mathrm {0.0608}} $ &
$\displaystyle \sim 3.9\ 10^{11} \overline{\rm B}_d $\\ 
\hline
& & & \\

$\displaystyle {\mathrm {\overline{\rm B} \ decay}} $ &
$\displaystyle {\mathrm { K^{*0}\ Decay}} $ & 
$\displaystyle {\mathrm {Final}} $ &
$\displaystyle {\mathrm {Number \ of}} $ \\

$\displaystyle {\mathrm {Mode}}  $ &
$\displaystyle {\mathrm {Mode} } $ &
$\displaystyle {\mathrm{State}} $ &
$\displaystyle {\mathrm{\overline{B} \ decays}} $ \\ 
\hline \hline
& & & \\

$\displaystyle  \overline{\rm B}_s\to {\mathrm K^{*0} \overline{\mathrm K}^{*0}}$ &
$\displaystyle {\mathrm {K^+\pi^-}}$ &
$\displaystyle {\mathrm {K^+\pi^-K^-\pi^+}} $ &
$\displaystyle \sim 4.9\ 10^5 $\\ 
$\displaystyle  \overline{\rm B}_d\to {\mathrm K^{*0} \overline{\mathrm K}^{*0}}$ &
$\displaystyle {\mathrm { K^+\pi^-}}$ &
$\displaystyle {\mathrm {K^+\pi^-K^-\pi^+}} $ &
$\displaystyle \sim 1.4\ 10^5 $\\ 

%\hline
\hline

\end{tabular}   $$

\label{tab:Bs_decays}
\caption{\small \label{tab:Bs_decays} The expected number of produced $\overline{\rm B}_{d,s}$ decays to the specific decay mode ${\mathrm K^{*0} \overline{\mathrm K}^{*0}}$ at FCC-ee at a center of mass energy of $m_Z$ over 4 years with 2 detectors. This number has to be multiplied by 2 when including $\rm B_{d,s}$ decays. The branching fractions of the PDG~\cite{pdg:1} have been used.}
\end{table}
%%%%%%%%%%%%%%%%%%%%% END TABLE %%%%%%%%%%%%%%%%%%%%%%%%%%%%%
\noindent Table~\ref{tab:Bs_decays} summarizes the expected number of produced  $\rm (K^+\pi^-)_{K^{*0}}(K^-\pi^+)_{\overline{K}^{*0}}$ final states from $\rm B_{d,s}$ decays at the Z-pole at FCC.

\vskip 10pt
\noindent Although small, the main source of background is expected to be of combinatorial origin. Inclusive Monte-Carlo samples of Z $\rightarrow b \bar{b}$ and Z $\rightarrow c \bar{c}$ events have been used to confirm this expectation, and to quantify the level of the combinatoric background. The generated samples consist of about $10^9$ $b \bar{b}$ events and about $0.5\cdot 10^{9}$ $c \bar{c}$ events produced with the PYTHIA~8.306 Monte-Carlo generator~\cite{Bierlich:2022pfr}. PYTHIA generates also signal events in this inclusive $b \bar{b}$ sample, however the branching fractions are different than the values of PDG. For $\BdKK$, PYTHIA uses $10^{-6}$, instead of $0.83\cdot 10^{-6}$ (PDG) and for $\BsKK$, PYTHIA uses $4\cdot 10^{-6}$ instead of $11.1\cdot 10^{-6}$ (PDG), i.e. a factor $\sim 2.8$ too low. This is not an issue since we are mainly interested in the evaluation of the background level. The generated events were passed through a fast simulation of the IDEA detector~\cite{fccee:3}, which provides resolutions corresponding to a detector similar to the one described in this Section above. The simulation is based on DELPHES~\cite{de_Favereau_2014}.
In particular, the simulation software that turns charged particles into simulated tracks relies on a full description of the geometry of the IDEA vertex detector and drift chamber. The software accounts for the finite detector resolution and for the multiple scattering in each tracker layer and determines  the (non diagonal) covariance matrix of the helix parameters that describe the trajectory of each charged particle. This matrix is then used to produce a smeared 5-parameters track, for each charged particle emitted within the angular acceptance of the tracker. Finally, the events were subsequently analysed within the FCCAnalyses framework~\cite{FCCAnalyses}.  \\

\begin{figure}[htb]
\vfill
\begin{center}

%\vskip 5cm
\includegraphics[width=0.45\textwidth]{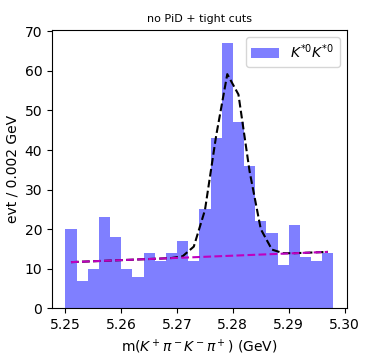}
\includegraphics[width=0.45\textwidth]{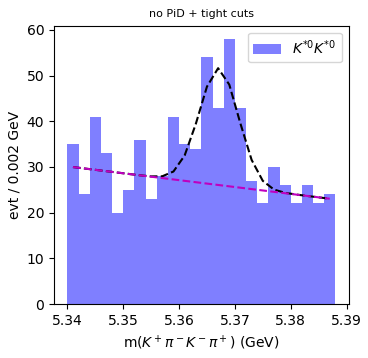}

\caption{\small \label{fig:signal_noPID} Invariant mass distribution for  $\BdKK$ and  $\BsKK$  candidates without particle identification.}
\end{center}
\vfill

\end{figure}

\noindent The reconstruction of signal candidates starts with the identification of the ``primary tracks'', that can be fit to a primary vertex\footnote{A simple iterative algorithm is used here. In a first step, all tracks are fit to a common vertex, using a constraint given by the beam-spot size. The track that gives the largest contribution to the $\chi^2$ of the fit is removed, and the remaining tracks are fit again. The procedure is repeated until the $\chi^2$ contribution of each track is below a given cut.}, and, consequently, of the ``secondary tracks''. Moreover, all reconstructed particles are used to determine the thrust axis, and the plane orthogonal to this axis and containing the interaction point divides each event in two hemispheres. \\
Quadruplets of secondary tracks with total charge 0 that belong to a same hemisphere are fit to a common vertex.  A $\chi^2  < 20$ is requested. Furthermore, pairs of particles with invariant mass (determined from the tracks' momenta at the fitted vertex) within $m_{K^{*0}}-0.075$ and $m_{K^{*0}}+0.075$ GeV define the $K^{*0}$ and $\overline{K}^{*0}$ candidates. The standalone vertex fit algorithm~\cite{BedeschiCode} used in this analysis is available in the distribution of the DELPHES package, and its recent extension to allow neutral particles to be included in the fit is described in~\cite{Franco_neutrals}. Quadruplets with an invariant mass above $5.0$ GeV and a momentum larger than 10 GeV are kept as potential $\rm B_s$ ($\rm B_d$) candidates. Table~\ref{tab:selection} summarizes all cuts.
%%%%%%%%%%%%%%%%%%%% Begin Table %%%%%%%%%%%%%%%%%%%%%%%%%%%%%%
\begin{table}[htb]

\centering
\small
$$ \begin{tabular}{lccccccccc}
\hline

$\displaystyle {cuts } $ &
$\displaystyle \rm {tr. 1-4} $ & 
$\displaystyle {\rm {K^{*0}}  cand} $ & 
$\displaystyle {\rm {\overline{K}^{*0}} cand} $ & 
$\displaystyle {\rm (K^{*0}\overline{K}^{*0}) cand}$ &
\\
\hline \hline

$\displaystyle {\rm |\cos\theta}|$ &
$\displaystyle {\rm <0.95 }$ &
$\displaystyle {\rm n/a}$ &
$\displaystyle {\rm n/a}$ &
$\displaystyle {\rm n/a}$ &
 
\\

$\displaystyle {\rm p\ (GeV)}$ &
$\displaystyle >0.5$ &
$\displaystyle {\rm n/a}$ &
$\displaystyle {\rm n/a}$ &
$\displaystyle >10.0$ &

\\
$\displaystyle {\rm m\ (GeV)}$ &
$\displaystyle {\rm PID}$ &
$\displaystyle {\rm m_{K^{*0}}\pm 0.075}$ &
$\displaystyle {\rm m_{K^{*0}}\pm 0.075}$ &
$\displaystyle {\rm >5.}$ &

\\
$\displaystyle \chi^2 \rm vtx$ &
$\displaystyle {\rm n/a}$ &
$\displaystyle {\rm n/a}$ &
$\displaystyle {\rm n/a}$ &
$\displaystyle 20$ &

\\ \hline

\end{tabular}   $$

\label{tab:selection}
\caption{\small \label{tab:selection} Summary of all cuts applied for selecting potential $\rm B_{d,s}$ candidates. The overall selection efficiency is about 33\%.}
\end{table}
%%%%%%%%%%%%%%%%%%%%% END TABLE %%%%%%%%%%%%%%%%%%%%%%%%%%%%%
\noindent We show in Figure ~\ref{fig:signal_noPID} the $\rm B_d$ and $\rm B_s$ candidates without particle identification. A signal is observed both for $\rm B_d$ and $\rm B_s$ but significant bakground is present. 
%%%%%%%%%%%%%%%%%%%% Figure Feynman diagram Bs to DsK  %%%%%%%%%%%%%%%%%%%
%%%%%%%%%%%%%%%%%%%%%%%%% End Figure %%%%%%%%%%%%%%%%%%%%%%%%%

%%%%%%%%%%%%%%%%%%%% Figure Feynman diagram Bs to DsK  %%%%%%%%%%%%%%%%%%%
\begin{figure}[htb]
\vfill
\begin{center}

%\vskip 5cm
\includegraphics[width=0.45\textwidth]{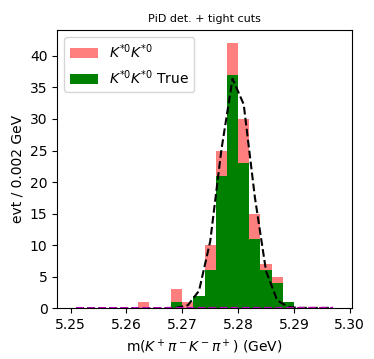}
\includegraphics[width=0.45\textwidth]{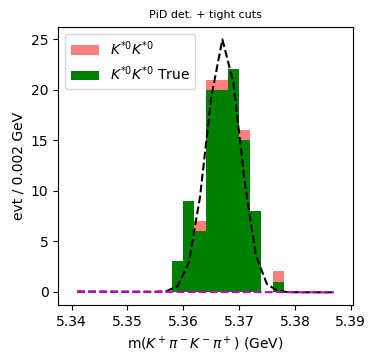}

\caption{\small \label{fig:signal_PID} Invariant mass distribution for  $\BdKK$ and  $\BsKK$  candidates with particle identification. The data in green are the genuine signal events while the data in red includes non-resonant backgrounds such as $\rm B_{d,s} \to  K^{*0} K^- \pi^+$, $\rm B_{d,s} \to  \overline{K}^{*0} K^+ \pi^-$ and at a lower level $\rm B_{d,s} \to  K^\mp\pi^\pm K^\pm \pi^\mp$ as well as the combinatorial background. }
\end{center}
\vfill

\end{figure}
%%%%%%%%%%%%%%%%%%%%%%%%% End Figure %%%%%%%%%%%%%%%%%%%%%%%%%
\vskip 10pt
\noindent Turning on the particle identification the situation changes dramatically (see  Figure ~\ref{fig:signal_PID}). As it can be seen, despite the low Branching Fractions, the signal is clearly visible with essentially no combinatorial background. However, one notes that there is some peaking background. These events are due to non-resonant (n-r) backgrounds such as $\rm B_{d,s} \to  K^{*0} (K^- \pi^+)_{n-r}$, $\rm B_{d,s} \to  \overline{K}^{*0} (K^+ \pi^-)_{n-r}$ and at a lower level $\rm B_{d,s} \to  (K^\mp\pi^\pm K^\pm \pi^\mp)_{n-r}$.
\subsection{A simple angular analysis}
\begin{figure}[htb]
\vfill
\begin{center}

%\vskip 5cm
\includegraphics[width=0.7\textwidth]{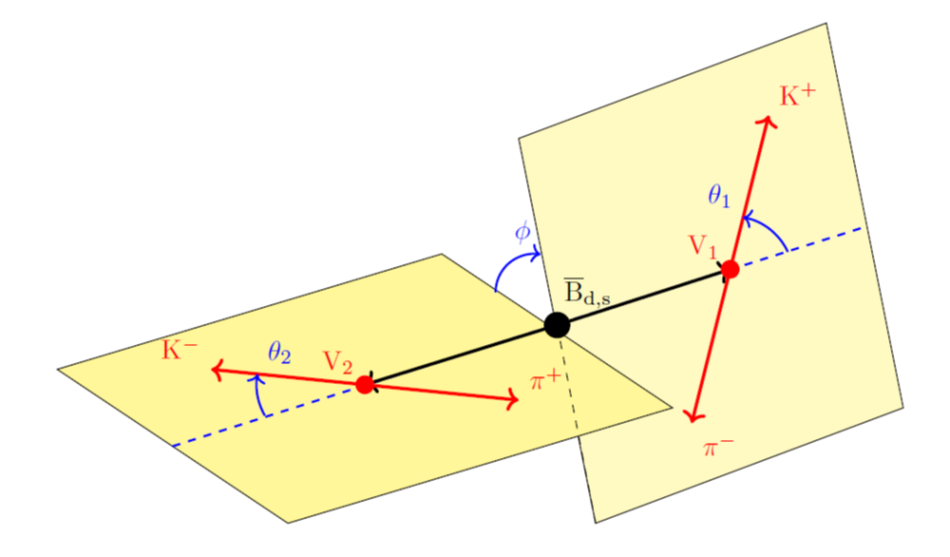}

\caption{\small \label{fig:schematic} Definition of the angles $\theta_1$, $\theta_2$ and $\phi$ for the decay $\rm \overline{B}^0\to  {\mathrm K^{*0} \overline{\mathrm K}^{*0}} $ used in the angular analysis. Each angle is defined in the rest frame of the decaying particle.}
\end{center}
\vfill

\end{figure}

\noindent Finally, we have carried out a simple angular analysis, i.e. without background, which we have shown to be small should one have an excellent PID, as also demonstrated by LHCb~\cite{lhcb:1}. To this end, we have generated $\overline{\rm B}_{d,s}$ decays to ${\mathrm K^{*0} \overline{\mathrm K}^{*0}}$ with the polarization expected with our calculations using QCD factorization, shown in Table~\ref{tab:B_decays_QCDF}. In pseudoscalar decays to 2 vector resonances decaying in turn to 2 pseudoscalar particles, the full angular dependence of the cascade reads as~\cite{KG:1} 
\begin{equation} 
\begin{array}{cccc} 
{d\Gamma (\overline{B}_{d,s}\to {\mathrm K^{*0} \overline{\mathrm K}^{*0}}) \over d\cos\theta_1 d\cos\theta_2 d\phi} \propto |{\cal \overline{A}}_0|^2 \cos^2\theta_1\cos^2\theta_2  +{ |{\cal \overline{A}}_+|^2 +  |{\cal \overline{A}}_-|^2 \over 4}\sin^2\theta_1\sin^2\theta_2 \\
\\
- [\Re (e^{-i\phi}{\cal \overline{A}}_0{\cal \overline{A}}^\ast_+)+\Re (e^{i\phi}{\cal \overline{A}}_0{\cal \overline{A}}^\ast_-)]\cos\theta_1\sin\theta_1\cos\theta_2\sin\theta_2 \\
\\
+ {\Re (e^{2i\phi}{\cal \overline{A}}_+	{\cal \overline{A}}^\ast_-)\over 2}\sin^2\theta_1\sin^2\theta_2
\end{array}
\label{19e} 
\end{equation}
\noindent where the angle $\phi$ is the angle between the decay planes of the two vector mesons in the B meson
rest frame and $\theta_{1,2}$ are the angles between the direction of motion of $\rm V_{1,2} \to P P$
pseudoscalar final states and the inverse direction of motion of the B meson as measured in
the $\rm V_{1,2}$ rest frame, see Figure~\ref{fig:schematic}. 
%%%%%%%%%%%%%%%%%%%% Figure Feynman diagram Bs to DsK  %%%%%%%%%%%%%%%%%%%
%%%%%%%%%%%%%%%%%%%%%%%%% End Figure %%%%%%%%%%%%%%%%%%%%%%%%%

%%%%%%%%%%%%%%%%%%%% Figure Feynman diagram Bs to DsK  %%%%%%%%%%%%%%%%%%%
\begin{figure}[hbt]
\vfill
\begin{center}

%\vskip 5cm
\includegraphics[width=1.1\textwidth]{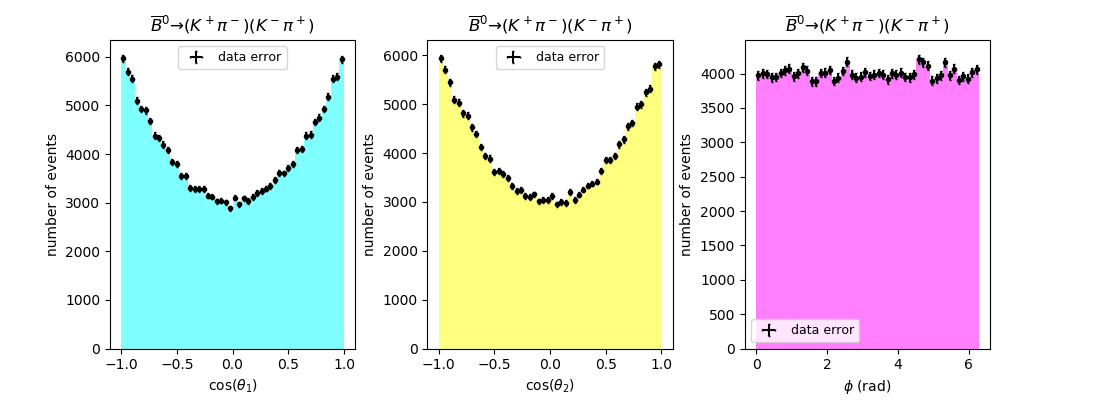}

\caption{\small \label{fig:polar_all} $\theta_1$, $\theta_2$ and $\phi$ distributions expected at FCC-ee for the decay $\rm \overline{B}^0\to  {\mathrm K^{*0} \overline{\mathrm K}^{*0}} $. The expected polarization fractions are $f_L=0.5$ and  $f_\parallel =f_\perp =0.25$.}
\end{center}
\vfill

\end{figure}
%%%%%%%%%%%%%%%%%%%%%%%%% End Figure %%%%%%%%%%%%%%%%%%%%%%%%%
\vskip 10pt
\noindent We show in Figure~\ref{fig:polar_all} the corresponding distributions for the $\rm \overline{B}^0\to{\mathrm K^{*0} \overline{\mathrm K}^{*0}}$ decay. Fitting the angular distributions with the dependences shown in equation~(\ref{19e}), one can extract the polarization fractions and estimate the statistical uncertainties.  Sensitivities at the level of few \textperthousand \ can be reached:
\begin{equation}
\begin{array}{cccc} 
\sigma_{f_{L,\parallel,\perp}}^{\rm B_d} & \simeq & 0.004 \\
\sigma_{f_{L,\parallel,\perp}}^{\rm B_s} & \simeq & 0.002 \\
\end{array}
\label{20e} 
\end{equation}

\noindent Such sensitivities would enable one to study in great detail the $\rm B$ decays into two vector states and maybe unravel New Physics.

% ================================ new section ===============================

\section{Conclusions}

In conclusion, we have made evident in a quantitative way that there is a problem of U-spin violation, very much larger than could be expected, in the decays $\BdsKK$, a trend that had been suggested by previous authors. This clear feature asks for new measurements, mainly on the decay $\BsKK$. For example, should the error on $f_{L, \mathrm{Exp}}$ for $\BsKK$ be reduced by a factor 2 in the near future, the significance would exceed 5 standard deviations. On a longer term, FCC-ee would enable to measure the polarizations with oustanding precisions for both ${\rm \overline{B}}_{d,s}$, allowing one to reveal whether new physics appears in $\BdsKK$. More generally, ${\rm \overline{B}}_{u,d,s}\to  V_1V_2$ is a very rich area for testing in depth the standard model further. 

{\subsection*{Acknowledgments}}
We wish to thank Franco Bedeschi for making his vertexing code available and for very useful discussions about the reconstruction of displaced vertices. We wish to thank also Emmanuel Perez for producing the $Z\to b\overline{b}$ and  $Z\to c\overline{c}$ simulation samples and for skimming the data. We are also indebted to Joaquim Matias for pointing out to us the interesting references~\cite{Alguero,Biswas}. \\

\end{document}